\documentclass{article}
\usepackage{amsmath}
\usepackage{amsfonts}
\usepackage{amssymb}
\usepackage{amsthm}
\usepackage{stackrel}
\usepackage{graphicx}
\usepackage{tikz}
\usepackage{xcolor}
\usepackage[export]{adjustbox}
\usetikzlibrary{arrows,positioning, shapes} 
\usepackage{float}
\usepackage{inputenc}
 \usepackage[T1]{fontenc}
\usepackage{url}
\usepackage{subfig}
\usepackage{amsthm,amsmath,amsfonts,amssymb}
\usepackage[numbers]{natbib}
\usepackage[colorlinks,citecolor=blue,urlcolor=blue]{hyperref}
\usepackage{graphicx}
\usepackage{float}
\usepackage{xcolor}
\usepackage{dsfont}
\usepackage{array}
\usepackage{color}
\usepackage{multirow}

\usepackage{float}

\theoremstyle{plain}

\newtheorem{theorem}{Theorem}[section]

\theoremstyle{definition}
\newtheorem{definition}[theorem]{Definition}
\newtheorem*{example}{Example}
\newtheorem*{remark}{Remark}

\title{Topological data analysis for random sets and its application in detecting outliers and goodness of fit testing}
\author{\footnote{ University of Split, Croatia  ( email: vgotovac@pmfst.hr)} Vesna Gotovac \DJ oga\v{s}, \footnote{University of Split, Croatia  ( email: mmandaric@pmfst.hr)} Marcela Mandari\'{c}}
\begin{document}
\maketitle
\begin{abstract}
   In this paper we present the methodology for detecting outliers and testing the goodness-of-fit of random sets using topological data analysis. We construct the filtration from  level sets of the signed distance function and consider various summary functions of the persistence diagram derived from the obtained persistence homology. The outliers are detected using functional depths for the summary functions. Global envelope tests using the summary statistics as test statistics were used to construct the goodness-of-fit test. The procedures were justified by a simulation study using germ-grain random set models.
\end{abstract}

\section{Introduction}

Random sets have gained popularity in recent years as a useful tool for statistically analysing the geometry of objects in various fields of science. A well-developed theory of random sets can be found in \cite{matheron:1975}, \cite{molchanov:2005} and \cite{serra:1982}. The strength of random sets lies in their ability to describe many phenomena in nature. Some of the many examples are research into the configuration of plants in the ecosystem (see e.g. \cite{moeller:2010}), in medicine to understand the properties of tissue (see e.g. \cite{hermann:2015}) or modelling the structure of the materials (see e.g. \cite{neumann:2016}).

For many problems, it is desirable to investigate the properties of random sets in terms of the shapes and the possible interactions between their components, such as clustering or repulsion tendencies. Many of these characteristics can be revealed by studying the topological properties of random sets realisations.
Since the methods of topological data analysis have proven successful in revealing the interactions and testing the goodness of fit in point processes (see \cite{tgfpp}), it seems reasonable to generalise these techniques for random sets.
This paper aims to pave the way for a methodology for detecting outliers and testing goodness of fit for random sets using topological data analysis.

Topology is a branch of mathematics that studies the shapes of objects based on their characteristics that are invariant under continuous deformations.
Topological data analysis (TDA) provides methods for analysing various topological features of given data sets.
In TDA, the data is transformed into a filtration, i.e. a nested family of objects that change with the given parameter and a persistent homology is used to track the persistence of the topological features in the filtration as the parameter changes.

Homology links a sequence of algebraic objects with other mathematical objects, in our case topological spaces. Homology groups were originally introduced to compare two shapes based on the basis of their $q$-dimensional holes. The 0-th homology group $H_0$ is associated with the connected components of the objects, the 1-th homology group $H_1$ with the loops or one-dimensional holes, and the 2-th homology group $H_2$ with voids or two-dimensional holes.
 Higher order homology groups can also be defined, but their interpretation is not as intuitive.

 Persistent homology captures the changes of these homology groups in the filtration as the parameter changes by tracking the births and deaths of $q$-dimensional holes. The features that persist over time are usually significant for describing the dataset.

 In the case of point processes, the filtration from which the persistent homology was obtained is constructed by assigning to each point a disc of increasing radius whose centre is the given point. As the radius increases, some components merge and die, also the holes may  appear and disappear as the radius continues to grow (for more details, see e.g. \cite{TDA}).

 The class of random set processes closely thigh to point processes are germ-grain models. In these models, the random set is formed as a union of simple random sets (grains) whose reference points (germs) form a point process.
It makes sense to try to generalise the point process method in the case of germ-grain models of random sets. The main difficulty in the generalisation is that, in contrast to the realisations of point processes, we do not observe the germs and grains in the realisation of germ-grain models, but only the union of the grains positioned at the germs.
If we construct a filtration on the realisation of germ-grain model by increasing the components by the growing radius, the information about the diameter of the components cannot be revealed.

Therefore, we propose to use the signed distance function of the realisation and form a filtration by considering its sublevel sets. It is expected that with this approach we will be able to distinguish better shapes occurring in realisations of random sets in terms of clustering or repulsion tendencies as well as the size of the clumps formed by the grains. In cases where the grains are not highly overlapping, this approach could also reveal their diameters.

%Since the signed distance function will be presented in a form of a pixelised image and the corresponding sublevel sets consists of unions of cubes (pixels) we will involve the cubical persistent homology \cite{cubical}.

Persistence diagrams can be used to graphically represent the structure of persistence homology.

Persistence diagrams (PD) consist of the points whose x-coordinate is the birth time and y-coordinate is the death time of a certain feature within the filtration.

\begin{figure}[ht!]
    \centering
    \includegraphics[scale=0.8]{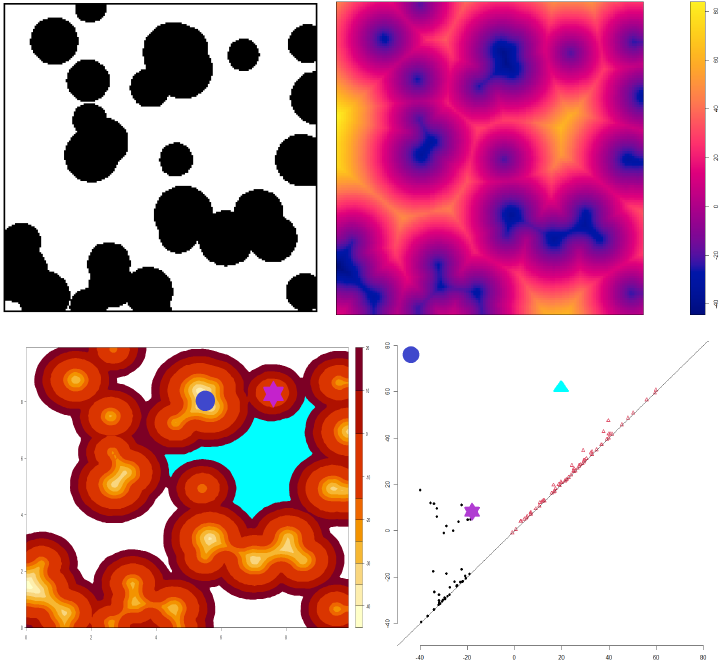}
    \caption{Top left: Realisation of the germ-grain model of the random set; top right: Signed distance function of the realisation of the random set; bottom left: sublevel set of the signed distance function for the thresholds -45,-40,-35,-30,-25,-20,-20,1,10 and 20; Bottom right: Persistence diagram corresponding to the sublevel sets of the signed distance function (black dots, purple star and blue dot represent the 0-dimensional homology and pink triangles with the blue triangle correspond to the 1-dimensional homology). }
    \label{fig:uvod}
\end{figure}

Let us illustrate our methodology on the  example depicted in Figure \ref{fig:uvod}. 
The first row of Figure \ref{fig:uvod} is the realisation of the germ-grain model (left image) and its signed distance function (right image). The left image in the second row of Figure \ref{fig:uvod} represents sublevel sets of the signed distance functions for the thresholds -45, -40, -35, -30, -25, -20, 1, 10 and 20. The blue dot represents the global minimum of the signed distance function, which is the first-born component. The component labeled with the purple star first appeared (was born) at -20 and died at sublevel 10 when it merged with the growing component that started to grow in the blue dot. Also note that the light blue colored hole was  born at sublevel 20 and will die at sublevel 60.

 The bottom right image in Figure \ref{fig:uvod} shows the PD obtained from the filtration of the signed distance function of the random set realisation from the same figure. The 0-dimensional persistent homology PD consists of black dots together with the purple star corresponding to the birth and death coordinates of the component labeled by the purple star in the bottom left image and the blue dot corresponding to the oldest connected component. The triangles correspond to the 1-dimensional PD, with the blue triangle representing the birth and death coordinates of the blue hole from the bottom left image.
 
  Since it is easier to work with functions in statistical analysis than with more complex objects such as measures, various summary functions of PD have been proposed. One of them is the accumulated persistence function (APF) (see \cite{APF}), which cumulatively sums the values of the rotated and rescaled PDs. 

  PD can also be regarded as an empirical measure.
  Since different PDs can produce the same APFs, we also propose a new way of uniquely representing  the PD as a function in terms of the support function of the lift zonotop of the empirical measure induced by the PD (see \cite{zonoids}).

Therefore, we will assign a function derived from its persistence diagram to each realisation of a random set.

To detect outliers in the sample of a random set, we will use the statistical depths. Statistical depths were first introduced to generalise the concepts of quantiles and ranks for multivariate data (see e.g. \cite{zuo:2000}). They rank the elements in the support of the distribution in the centre outwards, assigning values closer to 1 to the elements near the centre. The further you move away from the centre of the distribution, the smaller the depth values of the elements become and disappear in the case of elements that do not belong to the support of the distribution. Using the depth values to detect outliers is simple: the elements whose depth values are below a certain threshold are considered outliers.
There are many generalisations of multivariate depths for functional data (integrated depth \cite{fraiman:2001}, band depth \cite{lopez:2009}, infimal depth (\cite{mosler:2013}, \cite{gijbels:2015})). In this paper, we will focus on concepts of functional depth, where functions are classified based on their shape \cite{gijbels:2017}.
Many concepts of depth for the random sets have been proposed in \cite{outliers_anz} for the case of convex compact random sets and in the general case \cite{rs_depth}. However, these depths are suitable for random sets whose realisations are in the form of a single connected component. Therefore, they are not applicable for more complex random sets such as germ-grain models. %to recognise of the components.

Global envelope tests for testing hypotheses related to functional data were introduced in \cite{GET} and are very popular because they provide a graphical representation of the results that could reveal the possible reasons for rejecting the null hypothesis.
We will exploit  persistence homology to test the goodness of fit of some germ-grain models by using summary functions of the obtained PDs as test functions for the global envelope test and compare them with the results obtained when using some standard  random set summary functions as test functions, such as the capacity functional evaluated at squares (\cite{chiu:2013}, \cite{molchanov:2005}) and the extended empty space function (\cite{chiu:2013}).
%, support function of the lift zonoid  
%It seems like it is a very useful tool in recognizing whether clustering or repulsiveness between the grains occur and can be used for the classification of random sets. 

The remainder of the paper is organised as follows. Section \ref{sec:2} recalls the existing theory on random sets, which is necessary for understanding the paper. Section \ref{sec:3} introduces the TDA methods for random sets. There, we consider APFs as summary functions obtained from PDs and introduce the support function of the lift zonotop of the PD as a new summary function. Section \ref{sec:4} deals with the methodology for detecting outliers in the samples of random sets using the functional depth of the summary functions. Section \ref{sec:5} presents the statistical tests of the goodness of fit of random set models based on the summary functions and justifies them by a simulation study.
The paper is concluded by the Discussion section, where we summarise our results and briefly reflect on the direction of our future work. 

\section{Existing theory}
\label{sec:2}

\subsection{Support function and basic set operations} \label{sf}
Denote by $\mathcal{K}$ the family of all convex compact sets in $\mathbb{R}^d.$
The set $\mathcal{K}$ is equipped with Minkowski addition and
 multiplication with non-negative real numbers as follows:
\begin{itemize}
\item $\forall A,B \in \mathcal{K}, \ A \oplus B=\left\{u+v:u \in A, v \in B\right\},$
\item $\forall \lambda \in \mathbb{R}_+, \forall A\in \mathcal{K}, \ \lambda  A=\left\{\lambda u : u \in A\right\}.$
\end{itemize}

\begin{definition}
    Support function of a convex compact set $A\in \mathcal{K}$ is a function $h(A,\cdot):\mathbb{R}^d \to \mathbb{R}$  defined by
\[
h(A,u) = \sup_{x \in A} \langle u,x \rangle, \ u  \in \mathbb{R}^d.
\]
\end{definition}

\begin{example}
\label{ex:supp_seg}
    The support function of the segment in $\mathbb R^d$ with the endpoints $a$ and $b$ has the form$\max \{ \left\langle a,u\right\rangle, \left\langle b,u\right\rangle \}.$
\end{example}
Since $\langle u, \cdot \rangle$ is a continuous function and the set $A$ is compact, $h(A,u)$ is well-defined and the supremum is attained for all $u \in \mathbb{R}^d$. It is also easy to see that the support function is convex and therefore continuous.

The support function respects operations such as Minkowski addition and multiplication with non-negative real numbers, i.e. for $A,B \in \mathcal{K}$ and $\lambda \in \mathbb{R}_+$ we have
\begin{equation} \label{eq1}
\begin{aligned}
     h(A\oplus B,\cdot)=h(A,\cdot)+h(B,\cdot), \\
     h(\lambda A, u)=\lambda h(A, u).
\end{aligned}
\end{equation}

Moreover, because of (\ref{eq1}), the support function $h(A, \cdot)$ is uniquely determined by its values on $S^{d-1},$ where $S^{d-1}$ is the unit sphere in $\mathbb{R}^d.$ In the following, we assume that the domain of $h(A, \cdot)$ is $S^{d-1}.$

Consider $X \subseteq \mathbb R^2$ and $A$ a subset of $\mathbb R^2$ containing the origin.
Denote $\check{A}=\{-a:a\in A\}$ the set that is symmetric to $A$ with respect to the origin.
%\item Then
\begin{definition}
The dilation and erosion of the set $X$ by the compact structuring element $A$ are defined respectively as \cite[p.~40--47]{serra:1982}:
\begin{eqnarray*}
D_A(X)&=&X\oplus A=\{z \in \mathbb R^2:\check{A}\cap X\neq\emptyset\}=\bigcup_{a\in A}X_a,\\
E_{A}(X)&=&X\ominus A=\{z \in \mathbb R^2 :A_z\subseteq X\}=\bigcap_{a\in A}X_{-a},\\
%O_A(X)&=&D_A(E_A(X))=(X\ominus \check{A})\oplus A,\\
%C_A(X)&=&E_A(D_A(X))=(X\oplus A)\ominus \check{A},
\end{eqnarray*}
where $X_a$ is the set $X$ translated by $a$, i.e.~$X_a=\{x+a;x\in X\}$.
\end{definition}

For more details on morphological operations, see e.g. ~\cite{serra:1982,soille:2003}.

\subsection{Random sets}
The definitions in this section are taken from \cite{molchanov:2005} and \cite{chiu:2013}, unless otherwise stated.

Let
$\mathcal F$
 denote the family of closed sets in
$\mathbb{R}^d$
and
$\mathcal C$ the family of compact sets in $\mathbb R^d.$
\begin{definition}
\label{def:randomset}
 Let $(\Omega, \Sigma, P)$ be a probability space.
A mapping $\mathbb{X} :\Omega \to \mathcal{F}$ is a \emph{random closed set} if for every compact set $K \in \mathcal{C}$
$$\left\{ \omega \in \Omega : \mathbb X (\omega) \cap K \neq \emptyset \right\} \in \Sigma.$$
\end{definition}

\begin{definition}
\emph{The distribution $P_{\mathbb X}$  of  a  random  closed  set} $\mathbb X$  is  given  by the  relation $P_{\mathbb X}(F) =
P\left( \left\{
  \omega \in \Omega: \mathbb X(\omega) \in F
\right\} \right)$
for $F \in \mathcal{B}(\mathcal{F}),$ where $\mathcal{B}(\mathcal{F})$ is  the  Borel sigma algebra on $\mathcal F$ generated by Fell topology.
\end{definition}

%vidi jel \mathbb{X} i onda X(omega)
\begin{definition}
    A functional $T_{\mathbb{X}} \colon \mathcal{C} \to [0,1]$ given by $$T_{\mathbb{X}}(K)=P(\{\omega: \mathbb{X}(\omega) \cap K \neq \emptyset\}), \quad K \in \mathcal{C},$$ is said to be \textit{the capacity functional of random set} $\mathbb{X}$.
\end{definition}
The capacity functional uniquely determines the distribution of a random closed set \cite{molchanov:2005, chiu:2013}.
Since in practice it is not possible in practice to derive the capacity functional for all $K \in \mathcal C,$ we assume stationarity of the random set and consider the values of the capacity on the squares,
$T_{\mathbb X}(r)=T_{\mathbb X}(rB),$ where $B$ is a unit square.

%\begin{definition}
%    Function $F \colon \mathbb{R}^+ \to [0,1]$ given by $$F(r)=P(o \in \mathbb{X} \cap B(0,r)),$$ where $o$ stands for the origin is said to be  \emph{empty space function} of random set $\mathbb{X}$.
%\end{definition}
%Empty space function can be viewed as the distribution function of the random distance from the origin to the random set $\mathbb X.$

\begin{definition}
 Function $F \colon \mathbb{R} \to [0,1]$
 given by
$$
F(r)= \begin{cases}\mathrm{P}(o \in \mathbb X \oplus r {B}(o,1)) & \text { for } r \geq 0, \\ \mathrm{P}(o \in \mathbb X \ominus|r| B(o,1)) & \text { for } r<0 ,\end{cases}
$$
where $o$ stands for the origin and $B(o,1)$ for the unit disc centred at the origin, let  be \emph{ the extended empty space function} of the random set $\mathbb{X}$.
\end{definition}

The extended space function can be regarded as a distribution function of the signed distance from the random set $\mathbb X,$ i.e. the distance from the random point to the edge of the random set $\mathbb X$ with a positive sign if the randomly selected point doesn't fall within $\mathbb X$ and with a negative sign if the random point hits $\mathbb X$. 
\begin{definition}
    A random vector $\xi$ in $\mathbb{R}^d$ is a selection of $\mathbb{X}$ if $P(\xi \in \mathbb{X})=1$. The family of all integrable selections of $\mathbb{X}$ is denoted by $L^1(\mathbb{X})$.
\end{definition}
\begin{definition}
    A random closed set $\mathbb{X}$ is integrable if the family $L^1(\mathbb{X})$ is non-empty. For integrable random closed sets $\mathbb {X}$ selection expectation is defined by
\begin{equation}
\mathbb{E}(\mathbb{X})=\operatorname{cl}\left\{\mathbb{E}(\xi): \xi \in L^1(\mathbb{X})\right\}.
\label{eq:exp}
\end{equation}
\end{definition}
\begin{example}
\label{ex:fin_supp_exp}
 If $\mathbb X$ is a simple random convex compact set taking values $F_1, \ldots, F_k$ with probabilities $p_1, \ldots, p_k$, then the selection expectation of $\mathbb X$ is given by the weighted Minkowski sum
$$
\mathrm{E} \mathbb X =\operatorname{cl}\left(p_1 F_1\oplus \cdots \oplus p_k F_k\right).
$$
\end{example}

If $\mathbb X$ is a.s. convex compact then
$$
\mathbb{E} h(\mathbb X, u)=h(\mathbb{E} \mathbb X, u), \quad u \in \mathbb{S}^{d-1}.
$$

\begin{definition}[\cite{chiu:2013}]
\label{def:boolean}
Let $Y = \{y_1,y_2,\ldots\}$ be a stationary Poisson point process in
$\mathbb R^d$ and  $\{\mathbf B_1, \mathbf B_2, \ldots\}$ a sequence of
independent identically distributed random compact sets in $\mathbb R^d$
which are independent of $Y$. 
If $\mathbb E |\mathbf B_1 \oplus K| < \infty$ for all $K \in \mathbb K$, then the random set
\begin{equation}
\label{eq:defboolean}
\mathbb X = \bigcup_{n=1}^{\infty}(y_n + \mathbf B_n)
\end{equation}
is called \textit{Boolean model}.
The point process $Y$ is called the process of germs and $\{\mathbf B_1, \mathbf B_2, \ldots\}$ are referred as grains.
\end{definition}

The next step in the generalisation of the Boolean model of random sets is \textit{a germ-grain model} in which the Poisson point process is replaced by an arbitrary point process. The dependency between the germs and grains is also allowed.

\begin{definition}
\label{def:quermass}
 Let $\mathbb X$ be the Boolean model with 
$\mathbf B_1$ being a disc in $\mathbb R^2$ with the random radius.
\emph{Quermass-interaction process} is a random set 
whose probability measure 
is absolutely continuous with respect to the 
probability measure of $\mathbb B$ with density given by
\begin{align*}
	f_{\mathbf\theta}(\mathbf b)
    & = 
    \frac{1}{c_{\mathbf\theta}}
    \exp\{
    	\theta_1A(  U_{\mathbf b})
        + \theta_2L(  U_{\mathbf b})
        + \theta_3\chi(  U_{\mathbf b})
    \}
\end{align*}
for each finite disc configuration
$\mathbf b = \{\mathbf b_1\ldots,\mathbf b_n\}$, where
$A=A(  U_{\mathbf b})$ is the area,
$L=L(  U_{\mathbf b})$ is the perimeter,
$\chi=\chi(  U_{\mathbf b})$ is Euler-Poincar\'{e} characteristic
(i.e. the number of connected components minus the number of holes)
of the union $U_{\mathbf b} = \cup_{i=1}^n \mathbf b_i$, 
$\theta = (\theta_1,\theta_2,\theta_3)$ is 3-dimensional vector 
of parameters and $c_{\theta}$ is the normalising constant.
\end{definition}

\begin{figure}[H]
 \includegraphics[scale=0.65]{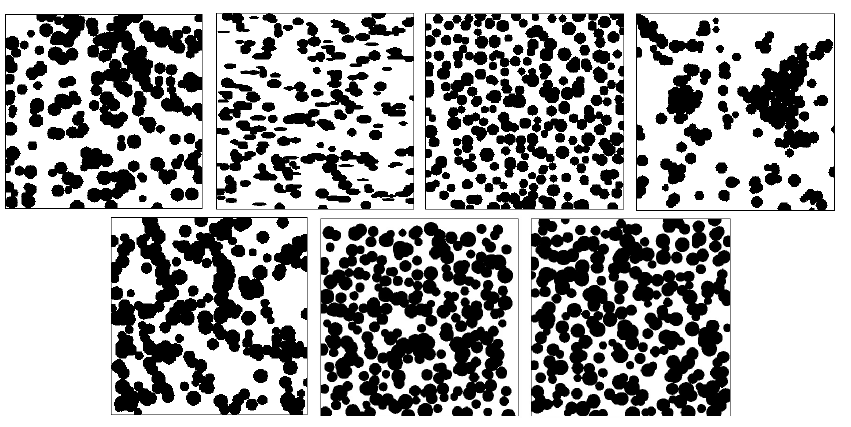}
 \caption{One realisation of each random set process in this order: First row: Boolean, Boolean ellipse, repulsive, cluster, Matern cluster, Cell and DPP}
\label{fig:procesi}
\end{figure}

\begin{figure}[H]
    \centering
    \includegraphics[scale=0.65]{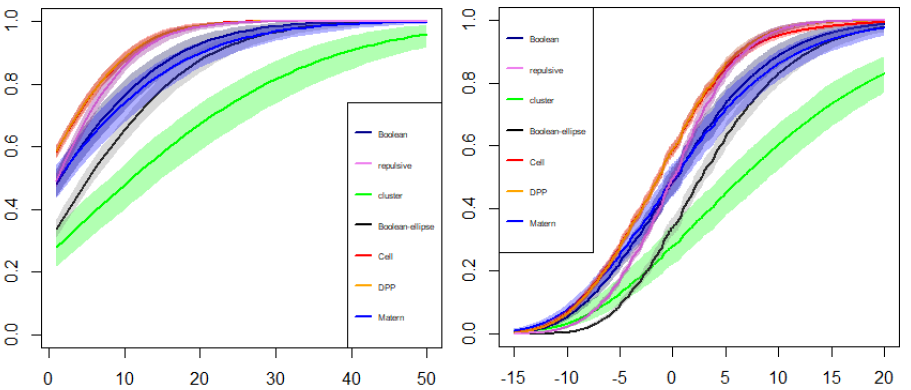}
    \caption{Mean capacity functional of each process at the left and mean extended empty space functions at the right together with their $95\%$ envelopes obtained by simulating 100 realisations of each random set process.}
    \label{fig:fk_es}
\end{figure}

In our simulation studies we will use the following random set models:

\begin{itemize}
 \item A Boolean model where $\mathbf B_1$ are discs with random radii (referenced as \emph{Boolean} model), where the centres of discs are in the window $25 \times 25$, the intensity of the disc centres is equal to $0.4$ and the radii are uniformly distributed on the interval $\langle 0.5, 1\rangle,$
 \item A Boolean model where $\mathbf B_1$ are ellipses with random major and minor axes (referenced as \emph{Boolean ellipse} model) with ellipse centres of ellipses in the window $25 \times 25$, the intensity of the ellipse centres equal to $0.4$ and uniform distribution of the semi-major axes on the interval $\langle 0.5, 1 \rangle$ and the semi-minor axes on the interval $\langle 0.2,0.7 \rangle.$
 \item the Quermass-interaction process with the parameters $\theta_1$ = 0.62, $\theta_2$ = -0.86 and $\theta_3$ = 0.7 with respect to the above-mentioned Boolean model (referenced as \emph{cluster} model, see \cite{kendall:1999}).
 \item the Quermass-interaction process with parameters $\theta_1$ = -1, $\theta_2$ = 1 and $\theta_3$ = 0 with respect to the same random-disc Boolean model (referenced as \emph{repulsive} model, see \cite{kendall:1999}).
 \item Model were the germs form a Matern cluster point process (see e.g. \cite{matern}) with intensity $0.4$ and the grains are discs with uniformly distributed radii on the interval $\langle 0.5, 1 \rangle$, which is referred to as \emph{Matern cluster} later in the text.
   \item Model with germs forming a Baddeley-Silverman cell process (see e.g. \cite{baddeley}) with an intensity of $0.4$ and the grains are discs whose radii are uniformly distributed on the interval $\langle 0.5, 1 \rangle$, which is referred to later in the text as \emph{Cell process.}
 \item Model where the germs are a determinant point process of Bessel type (see e.g. \cite{DPP}) with intensity $0.4$ and the grains are discs with radii uniformly distributed radii on the interval $\langle 0.5, 1 \rangle$, which is referred to as \emph{DPP} later in the text.

 Figure \ref{fig:procesi} shows a realisation of each of the mentioned random set models.
\end{itemize}
Figure \ref{fig:fk_es} shows the mean capacity functional and the extended empty space function for all models considered models together with their $95\%$ envelopes obtained using 100 simulations. 
%C:\Users\fesb\Documents\ESF i CF na jednom grafu.R
\section{Topological data analysis for random set realisations}

\label{sec:3}
The core of any TDA method is the construction of the filtration, i.e. a parameterised family of sets increasing with the parameter in the sense of inclusion, with respect to which the persistent homology is then obtained. In this section, we propose a way to construct the filtration based on random set realisations.

Suppose we observe the realisations of our random set within the observation window $W \subseteq \mathbb R^2.$
Let us consider a signed distance function $f_d: W \to \mathbb R$ as
\[f_d(x)=\left\{\begin{array}{ll}
d(x,S), & x \notin S,\\
-d(x,W\diagdown S) & x \in S,
\end{array}\right. \]
where $d(x,S)$ is the distance from a point $x$ to a set $S,$
and if we consider $S_r$  the sublevel sets of $f_d,$ i.e.
$$S_r=f_d^{-1}\left(\left\langle-\infty,r \right]\right).$$
In this way, we obtain a non-decreasing filtration $(S_r)_{r \in \mathbb{R}}$, from which we can construct a persistence diagram.

Without going into the theoretical details of constructing persistence diagrams, we can take a look at a simple example to understand what we can see in the persistence diagram.
\begin{example}
Consider the union of the discs in Figure \ref{fig:ex_balls}, whose centres lie on the circle around the origin with radius $200$. The radii of the discs are  \break $100,90,80,70,65,60,55,40,35,25$. On the heat map of the signed distance function $f_d$ it can be seen that the local minima are obtained near the centres of the spheres and the values correspond approximately equal to the negative value of the radius. The local maximum is reached near the centre of the "hole" enclosed by the discs.
The number of points in the persistence diagram for the $0$ dimension (black dots) corresponds to the number of discs, where the first coordinate is the negative value of their radius. It can also be seen that a point for the $1$ dimension (marked with red triangle) protrudes above the diagonal. It represents the hole enclosed by the discs.

We can imagine these discs as the silhouette of some trees and assume that the growth of the trees is linear in time and the same rate applies to all trees. Then the sublevel set $S_r$ could be interpreted as the silhouette of the trees at  time $r$, and the $0$-dimensional persistence diagram captures the birth of each tree and the time at which this tree first touches a neighbouring tree.
\begin{figure}[ht]
    \centering
    \includegraphics[scale=0.7]{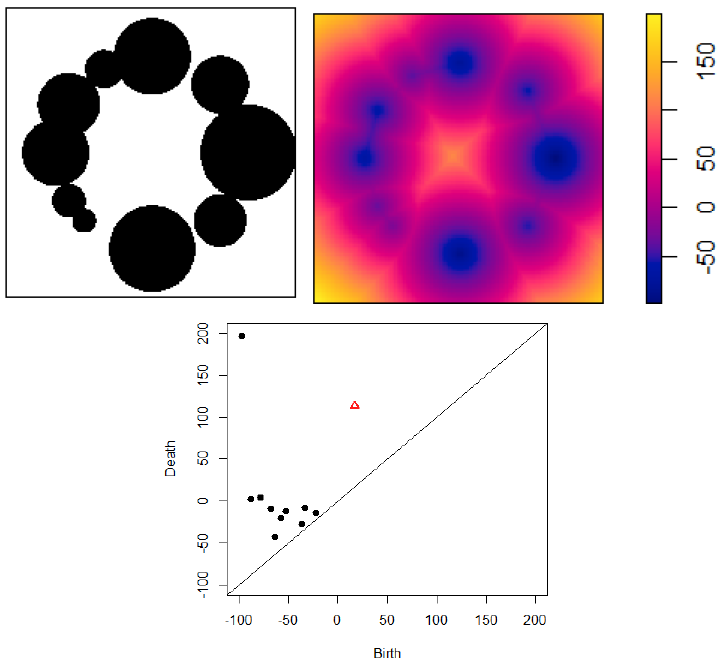}
    \caption{First row on the right is the heat map of signed distance function $f_d$ and persistence diagram for given example in the second row.}
    \label{fig:ex_balls}
\end{figure}
% C:\Users\fesb\Documents\Kapactitet\Primjer_kugle_pd.R
\end{example}

\begin{remark}
The alternative way to define a filtration on the realisation of a random set is to consider
\[S'_{+r}=(S \oplus r\check{B})\cap W,\]
where $\oplus$ stands for the Minkowski addition of sets and $B$ is an arbitrary structuring element, $r>0$ and $W$ denotes a given window that we observe.
Thus, we obtain a non-decreasing family of sets $(S'_r)_{r>0}.$ Tracing the birth and death of homological features using this filtration can reveal the types of clusters and the voids or connected components of the set. However, the information about the diameter of the components cannot be revealed.
So for an arbitrary $r>0$ we can consider sets
\[S'_{-r}=(S \ominus |r|B)\cap W,\]
where $\ominus$ stands for erosion.
Note that $(S'_r)_{r \in \mathbb R}$ is again a non-decreasing family of sets.
If the structuring element $B$ is equal to $B(0,1)$, the same filtration is obtained as when using the signed distance function.
Since the construction of the filtration based on the signed distance function was easier to implement, we didn't pursue this direction further.
\end{remark}

\begin{figure}[ht]
    \centering
   \includegraphics[scale=0.65]{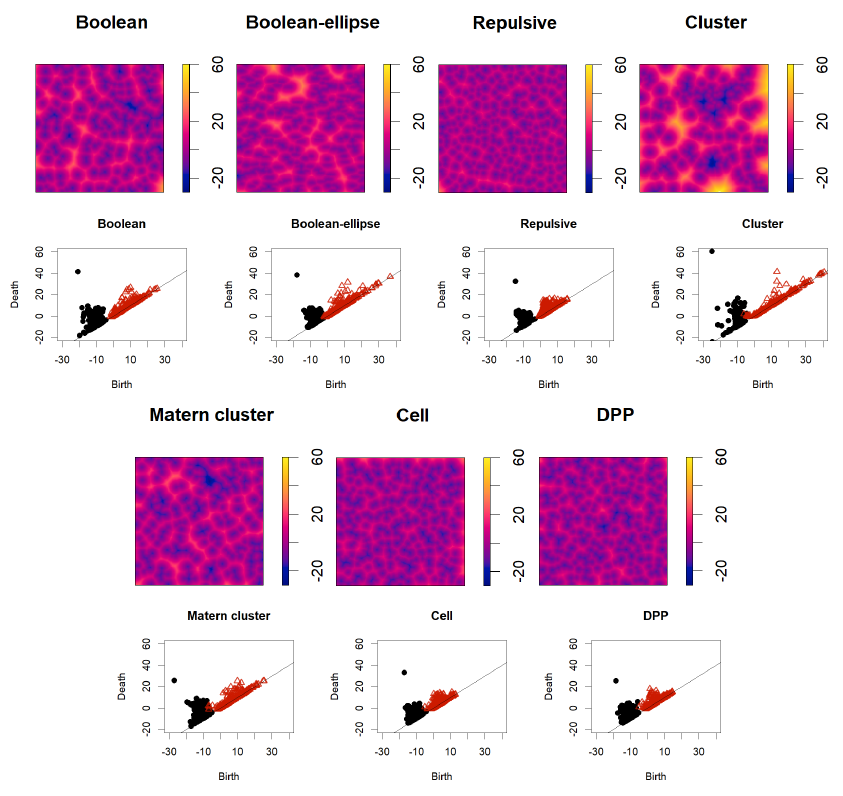}
    \caption{The heat maps of the signed distance function of one realisation of each process with its persistence diagram given in this order: Boolean, Boolean ellipse, Repulsive, Cluster, Matern cluster, Cell and DPP.}
    \label{proc1}
\end{figure}

Figure \ref{proc1} shows the heat maps of the signed distance function for a realisation of each  random set process mentioned in this paper together with its persistence diagram.
When calculating the signed distance function, the realisations were transformed into $400 \times 400$ matrices with values 0 and 1 corresponding to white and black pixels, respectively, and the function \verb{distmap{ from the package \verb{spatstat{ \cite{spatstat} in R was used. The PDs were obtained using function \verb{gridDiag{ from the package \verb{TDA{ \cite{TDA_R}.
% C:\Users\fesb\PD_sve_crtanje.R

When looking at the PD of the 0-dimension (black dots),
 the black dot with the smallest birth value (but the largest absolute birth value) and the largest death value represents the value range of the signed distance function. When comparing models with the same distribution of germs, this point usually has a larger absolute value for birth and death in cluster-tendency models such as the cluster and the Matern cluster models. They form components with larger diameter and have larger spaces, which results in the global extrema of the signed distance function being larger.
The cluster-tendency models also have a larger range of birth values of the point in the PD but a smaller number of points, while the repulsion-tendency models have a smaller range of birth values and a larger number of points in their PDs.

As for the 1-dimensional PD (red triangles in Figure \ref{proc1}), the models with clustering tendencies again exhibit a wider range of birth values when comparing the models with the same grain distribution, with the possibility of the birth of clusters with a negative sign (the hole forms in a negative sublevel of the signed distance function as a result of the overlap of the grains). Also in the case of the cluster model, which is characterised by the large empty areas, there are usually some points with high persistence that protrude from the diagonal and represent the large holes in the realisation. In the models with repulsion tendencies, whose realisations have a larger number of small holes, there is an overall larger number of points in the PD, but their birth and death areas are significantly smaller than in the models with cluster tendencies.

The persistence diagram can be regarded as an empirical measure
\[
PD^{q}(X)=\sum\limits_{i \in \mathcal I_q }\delta_{(b_i,d_i)},\] where $\mathcal I_q$ is an index set over all $q$-dimensional features and $b_i,$ $d_i$ are the birth and death times of the $i$-th feature, $q \in \{0,1\}$. Note that some features may have the same birth and death times, so the point $(b_i,d_i)$ may have a multiplicity $c_i$ greater than 1, which is not observable in the plots of PDs in Figure \ref{proc1}.

Our main goal is to use PDs as a tool for the statistical analysis of distributions of random sets.
Since in practice it is easier to deal with functions than with measures, we have considered various summary functions of PDs.
.

\subsection{Accumulated persistence function} 
We first consider accumulated persistence function introduced in \cite{APF}. For a given persistence diagram $PD^q,$ where $q$ is the dimension of the topological features it captures, we observe each point in the diagram as a triplet $(b_i,d_i,c_i)$, where $c_i$ denotes the multiplicity of each point, $b_i$ the birth time, and $d_i$ the death time.

\begin{figure}[H]
    \centering
   \includegraphics[scale=0.7]{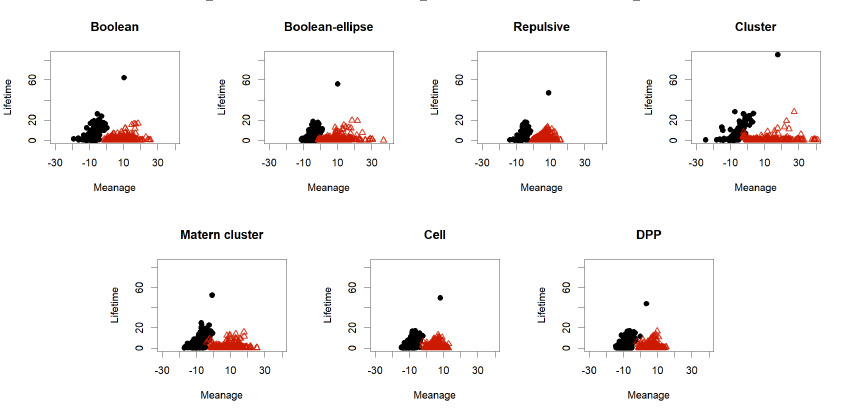}
    \caption{The RRPDs of realisations of randoms set processes in this order: Boolean, Boolean ellipse, Repulsive, Cluster, Matern cluster, Cell and DPP.}
    \label{RRPD}
\end{figure}

 We use $l_i=d_i-b_i$ to denote the lifetime and $m_i=\frac{b_i+d_i}{2}$ to denote the meanage of each feature in the diagram.
 We consider the so-called rotated and rescaled persistence diagram ($RRPD_q$), which is a diagram with points $(m_i, l_i, c_i),$ where $c_i$ is the multiplicity of each point.
 Figure \ref{RRPD} shows the $RRPDs$ of each process in the given order: Boolean, Boolean ellipse, repulsive, cluster, Matern cluster, Cell and DPP.
% C:\fesb\PD_sve_crtanje.R na kraju

\begin{figure}[ht]
    \centering
    \includegraphics[scale=0.65]{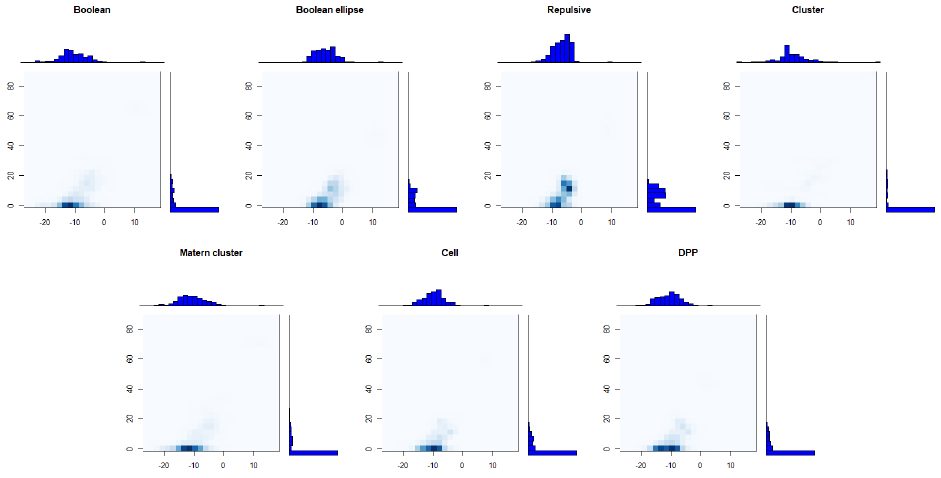}
    \caption{Rotated and rescaled persistence diagrams for dimension one as $2D$ histogram with regular $1D$ histograms bordering it, showing the density across each dimension. They are given in this order: Boolean, Boolean ellipse, Repulsive, Cluster, Matern cluster, Cell and DPP.}
    \label{rrpd}
\end{figure}

For a better understanding of what is seen on $RRPDs$ and what information it can give us about clustering or repulsion,
Figure \ref{rrpd} shows $RRPD_0$ for realisations of each process as a $2D$ histogram with regular $1D$ histograms on the edge showing the density across each dimension. Histogram gives  zhe distribution of meanages  on x-axes 
and the distribution of lifetimes on y-axes.

It can be seen from these histograms that processes with clustering tendencies ( cluster and Matern cluster ) have smaller meanages than those with repulsive tendencies (repulsive, Cell and DPP). This is because when clustering occurs, many connected components die quickly after their birth, which means that their meanage is smaller. Although most features die quickly after their death, there are a few components that persist for a long time and can be seen in the histograms as points with the largest meanages. These represent the largest clusters that die when they connect with another large component. In contrast, in the case of rejection, the deaths occur later, so that the meanages are larger. In the case of repulsion, lifetimes are also longer because there are more connected components that persist longer. Although all realisations have most lifetimes are around zero for all realisations, this is most evident when clustering occurs, which is due to the aforementioned "rapid" death of connected components.

Using the notation from the beginning of this subsection, we now give the definition of the accumulated persistence function ($APF$):
\begin{equation}
 APF_q(m)=\sum_{i=1}^n c_il_i \mathds{1}(m_i \leq m), \quad m \in \mathbb R,
\end{equation} where $\mathds{1} ( \cdot)$ is the indicator function and $q$ stands for the dimension of the topological features under consideration.
$APF_q$ cumulatively sums the lifetimes of the features with respect to their manage or equivalently $APF_q(m)$ sums the ordinates of all the points in the$RRPD_q$ with abscissas less than or equal to $m.$

Since our sets are random, the $RRPDs$ obtained are also random, so that the $APFs$ are  random functions.

$APF_q$ kepng a lot of information about the persistence diagram and by examining its shape we can understand more about the random set that generated the observed persistence diagram. For example, 

if $APF_0$ has many jumps and is large for small meanages $m,$ this may indicate clustering due to the early death of connected components. This leads us to the assumption that we can recognise whether or not clustering occurs in our random set by exploring its $APF_q.$ In our case, it only makes sense to explore the dimensions $0$ and $1$, i.e. $APF_0$ and $APF_1.$

\begin{figure}[H]
    \centering
    \includegraphics[scale=0.7]{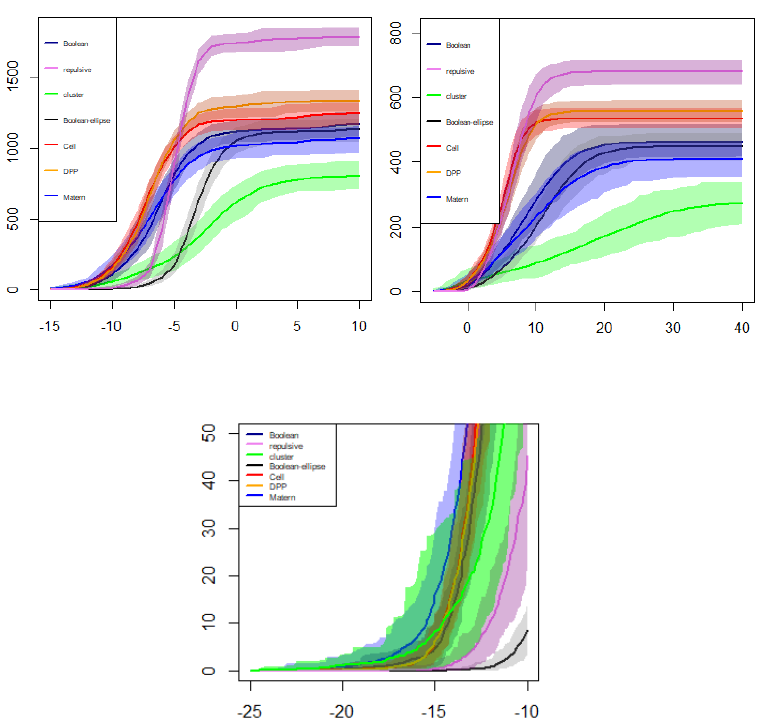}
    \caption{First figure shows mean values of $APF_0$ obtained from 100 simulated realisations of all observed processes with its $95\%$ envelope, second figure is closer look on $APF_0$ for small meanages $m$, third figure shows  mean values of $APF_1$ obtained from 100 simulated realisations of all observed processes with its $95\%$ envelope.}
    \label{4proc}
\end{figure}

Figure \ref{4proc} shows the mean values of $APF_0$ and $APF_1$ obtained from 100 simulated realisations of all considered models together with their $95\%$ envelopes. Note that $APF_0s$ shows more jumps and larger values for small meanages $m$ in the case of the cluster and Matern cluster models, which may indicate clustering.

When considering $APF_1s$, loops with longer lifetimes are born earlier in Boolean model realisations than in models in which clustering occurs, leading to a steeper increase in accumulated lifetimes. This steeper increase is even more pronounced in models where repulsion occurs. For small meanages $m$, especially below zero, models in which clustering occurs tend to have larger values of $APF_1$ as many small voids appear and disappear very early. However, this is short-lived as these values are soon surpassed by the Boolean, repulsive, Cell and DPP models due to the aforementioned steeper increase in accumulated lifetimes.

Taking everything into consideration, it seems that by exploring $APFs$ one can really find differences between realisations of random sets and give some sort of explanation for the differences that occur.

\subsection{Lift zonotop of a PD} \label{liftzonoid}
Since different PDs can generate the same APFs, we propose the use of the support function of the lift zonoid to uniquely describe  our PD with a function. 

Looking at the PD of the realisations of our random set models in Figure \ref{proc1}, the points with the highest persistence are few and there are many points with persistence close to zero (close to the line $x=y$). We also found that some of these points near the diagonal appear as computational errors. So the small percentage of the mass of $PD^{q}(X)$ is in the "important" points. Therefore, we propose to also consider the weighted empirical measure of the persistence diagram, which is defined as follows

\[
PD^{q}_W(X)=\sum\limits_{i \in \mathcal I_q }(d_i-b_i)\delta_{(b_i,d_i)},\] 
which has a mass $(d_i-b_i)c_i$ at the point $(b_i,d_i),$ where $c_i$ stands for the multiplicity of the point $(b_i,d_i)$ in the PD.

An integrable finite measure $\mu$
 on $\mathbb R^2$ is uniquely identified by a convex set $Z$ in $\mathbb R^3$ called a lift zonoid (see \cite{zonoids}). The set $Z$ is the expected value of the random segment in $\mathbb{R}^3$ with one endpoint at the origin and the other at $(1,\eta),$ where $\eta$ is distributed according to $\mu.$ (If $\mu$ is not a probability measure, the expectation is evaluated by replacing $\mathbb{E}(\xi)$ in (\ref{eq:exp}) by the integral of $\xi$ with respect to $\mu.$) 
 
 In our case, we will focus on the measure $PD^{q}_W(X)$, which has a finite support. According to Example \ref{ex:fin_supp_exp} its lift zonoid, called lift zonotop
 \begin{equation} \label{lz}
 Z=\bigoplus\limits_{i \in I_q}(d_i-b_i)\cdot[\mathbf{0},(1,b_i,d_i)],
 \end{equation}

i.e. the Minkowski sum of segments in $\mathbb R^3$ with one endpoint at the origin $\textbf 0$ and the other at the point $(1,b_i,d_i)$, multiplied by the scalar $l_i=(d_i-b_i)$, which represents the lifetime of the $i$th feature. Its support function $h_Z:S^2\to \mathbb R$ can be easily calculated using the properties (\ref{eq1}) and the Example \ref{ex:supp_seg} from Section \ref{sf} in the form
\begin{equation} \label{hlz}
    h_Z^q(u)=\sum\limits_{i \in \mathcal I_q} l_i\max{\{0,\left\langle u,(1,b_i,d_i)\right\rangle\}}, \ u \in S^2.
\end{equation}

Note that the value $h^q_Z(1,0,0)$ represents the sum of all lifetimes of the features (close to the number of local minima of the signed distance function or the number of connected components in the case of $q=0$, larger in the case of repulsion). The values $h^q_Z(0,-1,0)$ (for $q=0$ positively correlated with the sum of the diameters of the components) and $h^q_Z(0,1,0)$ (for $q=1$ dependent on the number of holes and their average diameter) are the weighted sum of all births that take place before or after time 0, respectively. $h^q_Z(0,0,-1)$ and $h^q_Z(0,0,1)$ (for $q=0$ positively correlated with the sum of the half-distances to the closest component) are the weighted sum of all deaths occurring before and after time 0, respectively.

\begin{figure}[H]
    \centering
    \includegraphics[scale=0.8]{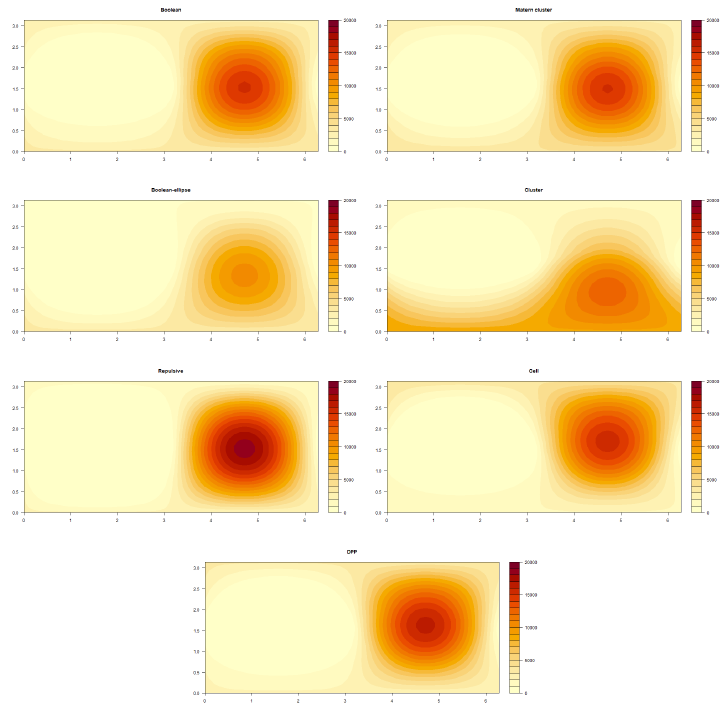}
    \caption{The support function of the lift zonotop of the measure associated with 0-dimension persistence diagram. They are given in this order: Boolean, Matern cluster, Boolean ellipse, Cluster, Repulsive, Cell and DPP.}
    \label{fig:hzonotop}
\end{figure}

\begin{figure}[H]
    \centering
    \includegraphics[scale=0.8]{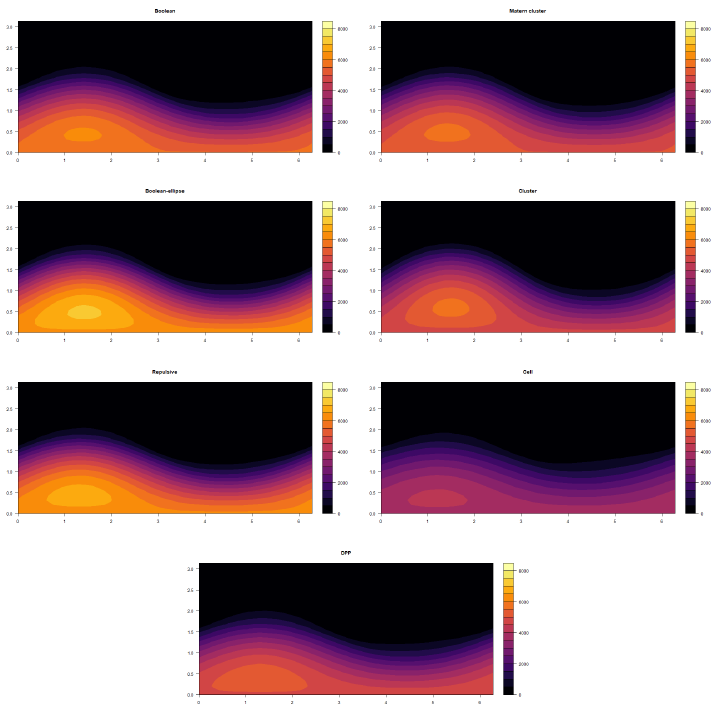}
    \caption{The support function of the lift zonotop of the measure associated with 1-dimension persistence diagram. They are given in this order: Boolean, Matern cluster, Boolean ellipse, Cluster, Repulsive, Cell and DPP.}
    \label{fig:hzonotop1}
\end{figure}

The Figures \ref{fig:hzonotop} and \ref{fig:hzonotop1} represent the support functions of the lift zonotopes of the measure associated with the measure $PD^{q}_W(X)$ for $q=0$ and $q=1,$ respectively. The range of the support function $S^2$ is parameterised in the usual way, i.e. we take $(\rho, \phi) \in [0,2\pi]\times[0,\pi]$ and identify $u \in S^2$ with $(\rho, \phi)$ such that $u=(\sin(\rho)\cos(\phi),\sin(\rho)\sin(\phi),\cos(\rho)).$

The figure \ref{fig:hzonotop} shows that the global maxima of the$h^0_Z,$ for almost all models except cluster, are obtained near $\rho=3\pi/2$ and $\phi=\pi/2$, which corresponds to $u=(0,-1,0)$ with the value $h^0_Z(0,-1,0).$

The repulsive model has the highest value because it has many components with a similar radius. The cluster model has the smallest values of $h^0_Z(0,-1,0)$ because its realisations usually have the smallest number of connected components: only a few components with large diameters and the others are the components with smaller diameters.
The global minima of $h^0_Z$ arise at $\rho=\pi/2$ and $\phi=\pi/2$, which corresponds to $u=(0,1,0)$ with the value $h_Z(0,1,0)=0,$ since there are usually no births after 0 in our case.

Figure \ref{fig:hzonotop1} reveals that the global maximum of $h^1_Z,$ is obtained near $\rho=\pi/6$ and $\phi=\pi/2$, which is $u=(\frac{\sqrt{3}}{2},\frac{1}{2},0)$ having value $h^1_Z(u)=\sum\limits_{i \in \mathcal I_1}l_i\max{\{0,\frac{1}{2}(\sqrt{3}+b_i)\}}$, which is usually highest for the Boolean ellipse process and the repulsive model with many holes that persist for a long time, followed by the cluster model with few holes and large $l_i$. Cell and DPP model have the smallest maximum values of $h^1_Z$ as their holes have the smallest lifetime.
The global minimum of $h^1_Z$ has the value 0 and it is obtained  e.g. for $\rho=\pi$ and $\phi=\pi/2$, which corresponds to $u=(0,0,-1)$, since the death of the clusters normally does not take place before time 0.

\begin{figure}[H]
    \centering
    \includegraphics[scale=0.65]{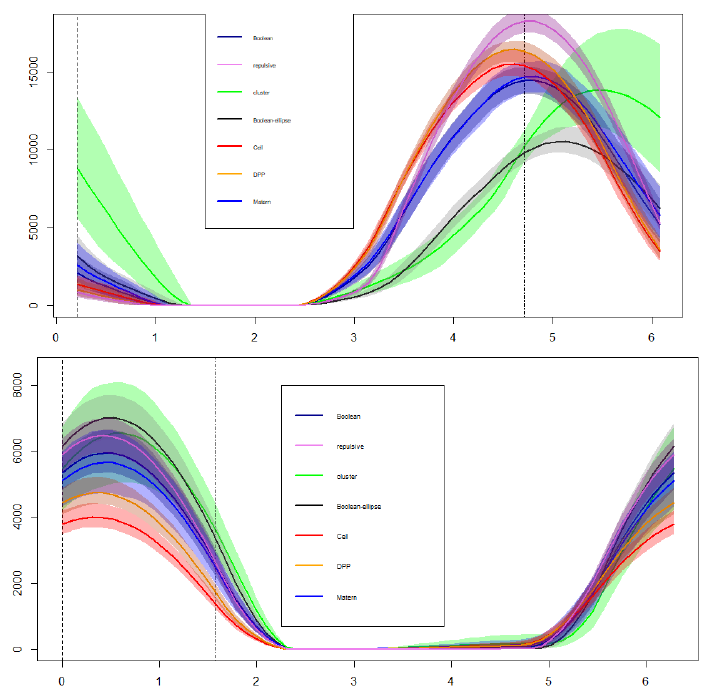}

    \caption{Top: Graphs of the average value of $h^0_Z$ for $\rho \in [0,2\pi]$ and fixed $\phi=\frac{\pi}{2}$ with respect to 100 realisations of each random set process with corresponding 95\% envelopes. The black dashed line represents the values for $u=(0,0,1)$, while the dot-dashed line represents the values for $u=(0,-1,0)$; below: Graphs of the average value of $h^1_Z$ for $\rho \in [0,2\pi]$ and fixed $\phi=\frac{\pi}{2}$ with respect to 100 realisations of each random set process with corresponding 95\% envelopes. The black dashed line represents the values for $u=(0,0,1)$, while the dot-dashed line represents the values for $u=(0,1,0).$}
    \label{fig:h_z_cross}
\end{figure}

Figure \ref{fig:h_z_cross} shows a visual representation of the mean support functions of lift zonoids for dimension $0$ and dimension $1$ obtained from $100$ realisations of each random set process together with $95\%$ envelopes when $\phi$ is fixed to the value $\frac{\pi}{2}$.

The upper image in Figure \ref{fig:h_z_cross} shows that cluster and Matern Cluster models have larger values for $h^0_Z(0,0,1)$, which suggests larger empty spaces. The repulsive, Cell and DPP models have larger values for $h^0_Z(0,-1,0)$, indicating a larger number of components or a larger average value of component diameters.

The bottom image in Figure \ref{fig:h_z_cross} suggests that $h^1_Z$ might be the best summary function to distinguish DDP and Cell models, since only in this case we have arguments whose $95\%$ envelopes do not overlap. The values of $h^1_Z(0,0,1)$ and $h_Z^1(0,1,0)$ are significantly smaller in the case of the Cell model compared to the DPP model, since the realisations of the Cell model have slightly smaller empty space areas on average than the DPP model, so the weighted sum of its births and deaths is smaller.

% C:\Users\fesb\SZ_crtanje_heatmapa.R

%Prva slika C:\Users\fesb\SZ0_svi_procesi_na jednoj slici.R
%Druga slika C:\Users\fesb\SZ1_svi_proceni_na jednoj slici.R

\section{Depth of realisations of random sets via random persistence diagrams} 
\label{sec:4}
To detect outliers in the sample of random sets, we use statistical depth. We will use the concept of functional depth as we identify each realisation of a random set from the sample with its corresponding summary function. 
%They provides us with ordering of the realisations with respect to the empirical distribution of the sample.

%to each persistence diagram and then analyse and compare depths of those functions.
%\subsection{Concept of statistical depth}

%Since there is no ordering in $\mathbb{R}^d$, 
Statistical depth is introduced as a way to generalise ordering and ranks that are well known in $\mathbb{R}$ \cite{zuo:2000}. It provides a centre outward ordering of elements in the support of the distribution. It is useful when searching for outliers or when comparing distributions \cite{liu:1990}. Realisations whose depth is below a certain threshold can be considered outliers, as this means that they are in some way "far" from the others in the sample.

In the multivariate case, the Tukey half-space depth \cite{Tukey} was introduced first, where the depth $D(x; P)$ of the point $x \in \mathbb R^d$ in a distribution $P$ on $\mathbb R^d$ is the minimum probability of the half-space containing the given point. Its empirical version is equal to the minimum of the percentages of points that lie in the half-spaces with the boundary containing the given point.

%For details we refer the reader to \cite{outliers_anz} where they can find many approaches to assigning depth to random sets.

%Sets can be represented as functions (see e.g. \cite{molchanov:2005}) in many ways, which can be very useful. It can be done for example by using an indicator function. Also, if the sets are convex bodies, then the most natural representation is by their support functions. 

%Since we will use this approach (meaning that we will assign a function to representation of random set), we will use the concept of functional depth.

%For example, the band depth (\cite{lopez:2009}) is defined as the probability of function being sandwiched between pointwise minima and maxima of fixed number of i.i.d. random curves and 

One of the possible generalisations for the functional domain is the so-called integrated depth (see \cite{fraiman:2001}). It is calculated as the integral over the univariate depths of the function values for the fixed arguments. We will focus on the functional depth approach, where the functions are classified based on their shape \cite{gijbels:2017}. In detail, the  $J$-th  order integrated depth of the function $f$ with respect to the random function $F$ with distribution $P$ is obtained as
$$
F D_J(f ; P)=\int_0^1 \cdots \int_0^1 D\left(\left(f\left(t_1\right), \ldots, f\left(t_J\right)\right)^{\top} ; P_{\left(F\left(t_1\right), \ldots, F\left(t_J\right)\right)^{\top}}\right) \mathrm{d} t_J \ldots \mathrm{d} t_1 .
$$
for $J=1,2,3.$

The 1st order outliers are the curves that differ from the pattern in terms of the values of the function, i.e. the minimum and maximum values. The 2nd order outliers are the curves that violate the pattern in terms of growth. The 3rd order outliers are the outliers in terms of convexity and concavity.

Our plan is to associate to each realisation of the random set the summary function: the capacity functional estimated on squares (CF), the extended empty space function (ESF), the APFs for dimensions 0 and 1 ($APF_0$ and $APF_1$), and the support functions of the lift zonotops of PD for dimensions 0 and 1 ($h_Z^0$ and $h_Z^1$). Then we calculate the depths of these summary functions with respect to the empirical distribution of the given sample of the functions. Based on the depths obtained,  we can decide which realisations can be considered as outliers.

 To calculate the depths, we use the function \verb{shape.fd.outliers{ from the package \verb{ddalpha{ in R, which implements the above-mentioned methodology for detecting  outliers in the sample of functions (see \cite{fd_outliers}). It returns which functions in the sample can be considered as the $J$-th order outliers for $J=1,2,3$.
 The function is classified as a first-order outlier if its 1st order integrated depth is within $\alpha \cdot 100\%$ of the smallest depth for a selected threshold value $\alpha \in \left\langle 0, 1 \right\rangle$. The extension of the boxplot method is used in the detection of 2nd and 3rd order outliers (for more details see \cite{gijbels:2017}).

%, we propose the use of so-called lift zonoid and its support function. In this subsection we state what is lift zonoid of a measure and how we find its support function. 

%Last step is finding support functions of obtained lift zonoids and

%\subsection{Simulation study}
To check how well the proposed methods work, we performed the simulation study with the realisations of the Boolean, repulsive, cluster, Boolean-ellipse, Matern cluster, Cell and DPP random set models.

%When simulating from the models, all the realisations are transformed to the matrices of 400 $\times$ 400 black and white pixels.

To see how well different summary functions recognise different models as outliers, we used the following approach.

We fix a model, simulate a sample of size 30 from that model, and want to see how well the methodology detects an additional realisation that comes from a different or the same model as an outlier within the simulated sample.
For each of the above models, we simulate the sample of 30 1000 times and add an additional realisation. We observe the percentage of cases in which the "intruder" (i.e., the one added realisation) was detected as a $J$th order outlier for any $J=1,2,3$ (a realisation can fall into exactly one category of $J$th order outliers). For the 1st order outliers, the threshold value $\alpha$ was set to $0.05$

Tables \ref{tab:B_dubine}, \ref{tab:R_dubine}, \ref{tab:C_dubine}, \ref{tab:Be_dubine}, \ref{tab:M_dubine}, \ref{tab:Ce_dubine}, \ref{tab:DPP_dubine} show the percentages of cases in which the one additionally added realisation was recognised as an outlier (i.e. it is $J$th order outlier for some $J=1,2,3$), if the simulated sample originates from the Boolean (B), the repulsive (R), the cluster (C), the Boolean-ellipse (Be), the Matern cluster (M), the Cell or the DPP for the additionally added realisations from all considered models.
The row names of the tables correspond to the various summary functions used,
%SZ0 and SZ1 for the support function of the lift zonotop of PD for dimension 0 and 1, C for the capacity functional, ESF for the extended empty space function, 
while the column names reveal the model of additional realisation. The percentages in bold are the highest in the given column and indicate which summary function differentiates the best model of additional realisation among the fixed models.

 %For calculating the depths we used the methodology from \cite{fd_outliers} (function \verb{shape.fd.outliers{  form package \verb{ddalpha{ in R).

 %repulsive, cluster and Boolean ellipse realisations were detected as the outliers. The cluster realisation was a first order outlier, while the repulsive and Boolean ellipse realisations were classified as the third order outliers. 

From Table \ref{tab:B_dubine}, we can conclude that the PD-based summary functions based on the PD were the most successful in detecting outliers in the Boolean model, except in the case of the Matern cluster model, where the capacity functional is the winner.
The $h_Z^0$ functions were best for detecting outliers in the Boolean ellipse and Cell models, while the APF0 and APF1 functions outperformed the other summary functions in detecting  outliers in the repulsive and DPP models.

Detailed percentages of outlier types (1st, 2nd and 3rd order) for all model combinations and added realisations can be found in Tables \ref{tab:out_type_B}, \ref{tab:out_type_R}, \ref{tab:out_type_C}, \ref{tab:out_type_Be}, \ref{tab:out_type_M}, \ref{tab:out_type_Cell} and \ref{tab:out_type_DPP} in Supplement \ref{sup}, if the simulated sample of 30 realisations originates from the Boolean, repulsive, cluster, Boolean-ellipse, Cell or DDP model.
Most of the realisations detected as outliers are 1st order outliers, except in the cases where the simulated sample model and the added realisation model have similar summary functions, then we have a significant percentage of  2nd and 3rd order outliers.

The given simulation study has shown that this method can be used to detect outliers in samples, as it detects the "intruders" well under the given random set, especially when using the summary functions based on PDs.

\begin{table}[H]
\caption{Percentages of detection of an additional model realisation from column names detected as outliers within the sample of 30 realisations of the Boolean model when different summary functions (row names) were used.}
    \label{tab:B_dubine}
    \centering
    \begin{tabular}{|c|c|c|c|c|c|c|c|}
    \hline
        & B & R & C &  Be & M & Cell & DPP \\ \hline
        $h_Z^0$ & 4\% & 97\%  & \textbf{100\%}  &  \textbf{97\%}  &  5\%  & \textbf{91\%}  &  92\% \\
        $h_Z^1$ & 9\%  & 60\% & 83\% & 65\%  &31\% & 77\% & 59\%\\
        $APF_0$ & 4\% & \textbf{100\%} &  \textbf{100\%} &  34\% & 67\% &  47\% &  \textbf{93\%}\\
        $APF_1$ & 10\% & \textbf{100\%} & \textbf{100\%} &  30\% & 42\% &  81\% &  \textbf{93\%}\\
        CF & 6\%  & 76\% & \textbf{100\%} &  79\% & \textbf{78\%} &  79\% & 75\%\\
        ESF &6\%  & 17\% & \textbf{100\%} &  52\% & 36\% &  8\% & 39\%\\
        \hline
    \end{tabular}
\end{table}

\begin{table}[H]
    \caption{Percentages of detection of an additional model realisation from column names detected as outliers within the sample of 30 realisations of the repulsive model when different summary functions (row names) were used.}
    \label{tab:R_dubine}
    \centering
    \begin{tabular}{|c|c|c|c|c|c|c|c|}
    \hline
        & B & R & C &  Be & M & Cell & DPP \\ \hline
        $h_Z^0$ & 99\% & 4\% & \textbf{100\%} & \textbf{100\%} & \textbf{100\%} & \textbf{100\%} &  99\%\\
        $h_Z^1$ & 82\% & 9\% & 97\% & 64\% & 92\% &\textbf{100\%} & \textbf{100\%}\\
        $APF_0$ & \textbf{100\%} & 9\% & \textbf{100\%} & \textbf{100\%} & \textbf{100\%} & \textbf{100\%} & \textbf{100\%}\\
        $APF_1$ & \textbf{100\%} & 15\% & \textbf{100\%} & \textbf{100\%} & \textbf{100\%} & \textbf{100\%} & \textbf{100\%}\\
        CF &71\% &  6\% & \textbf{100\%} &  77\% &  81\%&  79\%&  79\%\\
        ESF &99\% &  9\% & \textbf{100\%} & \textbf{100\%} & \textbf{100\%} & 95\% & 18\%\\
        \hline
    \end{tabular}
\end{table}

\begin{table}[H]
    \caption{Percentages of detection of an additional model realisation from column names detected as outliers within the sample of 30 realisations of the cluster model when different summary functions (row names) were used.}
    \label{tab:C_dubine}
    \centering
    \begin{tabular}{|c|c|c|c|c|c|c|c|}
    \hline
        &  R & B & C &  Be & M & Cell & DPP \\ \hline
         $h_Z^0$  & \textbf{100\%} & \textbf{100\%} & 10\% &\textbf{100\%} &\textbf{100\%} &\textbf{100\%} & \textbf{100\%}\\
        $h_Z^1$  & 64\% & 12\% & 11\% & 43\% & 14\% & 76\% & 40\%\\
        $APF_0$ & \textbf{100\%} & \textbf{100\%} & 11\% & \textbf{100\%} & \textbf{100\%} & \textbf{100\%} & \textbf{100\%}\\
        $APF_1$  & \textbf{100\%} & \textbf{100\%} & 6\% & \textbf{100\%} & 98\% & \textbf{100\%} & \textbf{100\%}\\
        CF  &  73\% & \textbf{100\%} &9\% &  78\% & 80\% &  81\% &  81\%\\
        ESF  & \textbf{100\%} & \textbf{100\%} &10\% &\textbf{100\%} &\textbf{100\%} &\textbf{100\%}& \textbf{100\%}\\
        \hline
    \end{tabular}
\end{table}

\begin{table}[H]
    \caption{Percentages of detection of an additional model realisation from column names detected as outliers within the sample of 30 realisations of the Boolean-ellipse model when different summary functions (row names) were used.}
    \label{tab:Be_dubine}
    \centering
    \begin{tabular}{|c|c|c|c|c|c|c|c|}
    \hline
        &  R & B & C & Be & M & Cell & DPP \\ \hline
$h_Z^0$ &    \textbf{95\%} &  \textbf{100\%}  & \textbf{100\%}  &   6\% & \textbf{100\%} & \textbf{100\%} &  \textbf{100\%}\\
$h_Z^1$  &  83\% &  36\%  & 96\%  &  10\%  & 88\% & \textbf{100\%} & \textbf{100\%}\\
$APF_0$    & 71\%  & \textbf{100\%} & \textbf{100\%} & 11\%   & 92\%  & 92\% & \textbf{100\%}\\
$APF_1$   & 49\%   & \textbf{100\%} & \textbf{100\%} & 11\%   & 71\% & \textbf{100\%} & \textbf{100\%} \\
CF    & 75\%  & \textbf{100\%} &  83\%   &  9\%&   80\% &  81\%  & 82\%\\
ESF     & 37\%  & 82\% & \textbf{100\%} & 11\% & 31\% &  68\%   & 86\%\\
        \hline
    \end{tabular}
\end{table}

\begin{table}[H]
    \caption{Percentages of detection of an additional model realisation from column names detected as outliers within the sample of 30 realisations of the Matern cluster model when different summary functions (row names) were used.}
    \label{tab:M_dubine}
    \centering
    \begin{tabular}{|c|c|c|c|c|c|c|c|}
    \hline
        &  R & B & C & Be & M & Cell & DPP \\ \hline
       $h_Z^0$  & 6\% & \textbf{100\%} & \textbf{100\%} & \textbf{87\%} & 7\% & \textbf{100\%} & 87\%\\
$h_Z^1$ & 2\% & 75\% & 53\% & 83\% & 7\%  & 50\% & 63\%\\
$APF_0$ & 58\% & \textbf{100\%} & \textbf{100\%} & 58\% & 18\%  & \textbf{100\%} & 93\%\\
$APF_1$ & 39\% & \textbf{100\%} & 99\% & 19\% & 8\%  & 99\% & \textbf{100\%}\\
CF & \textbf{73}\% & \textbf{100\%} & 85\% & 84\%  & 11\%  & 86\% & 83\%\\
ESF  & 5\% & 45\% & \textbf{100\%} & 18\% & 11\% &\textbf{100\%} & 31\%\\
        \hline
    \end{tabular}
\end{table}

\begin{table}[H]
    \caption{Percentages of detection of an additional model realisation from column names detected as outliers within the sample of 30 realisations of the Cell model when different summary functions (row names) were used.}
    \label{tab:Ce_dubine}
    \centering
    \begin{tabular}{|c|c|c|c|c|c|c|c|}
    \hline
        &  R & B & C & Be & M & Cell &DPP \\ \hline
$h_Z^0$ &  \textbf{95\%} & \textbf{100\%} & \textbf{100\%} & \textbf{100\%} & \textbf{100\%} & 2\% & 94\%\\
$h_Z^1$ &  83\% & \textbf{100\%} & 94\% & \textbf{100\%} & 93\% & 10\% & 84\%\\
$APF_0$ &  54\% & \textbf{100\%} & \textbf{100\%} & 80\% & \textbf{100\%} &  13\% &97\%\\
$APF_1$ &  93\% & \textbf{100\%} & \textbf{100\%} & 97\% & \textbf{100\%} & 9\% & \textbf{100\%}\\
CF &  74\% & \textbf{100\%} & 82\% & 79\% & 80\% & 10\% & 83\%\\
ESF &  72\% & 26\% & \textbf{100\%} & 92\% & \textbf{100\%} &7\% & 82\%\\

        \hline
    \end{tabular}
\end{table}

\begin{table}[H]
    \caption{Percentages of detection of an additional model realisation from column names detected as outliers within the sample of 30 realisations of the DPP model when different summary functions (row names) were used.}
    \label{tab:DPP_dubine}
    \centering
    \begin{tabular}{|c|c|c|c|c|c|c|c|}
    \hline
        & R & B & C & Be & M & Cell & DPP  \\ \hline
       $h_Z^0$  & 97\% & \textbf{100\%} & \textbf{100\%} & \textbf{100\%} & \textbf{100\%} & 97\% & 2\%\\
$h_Z^1$  & 83\% & \textbf{100\%} & 99\% & \textbf{100\%} & 95\% & 68\% & 6\%\\
$APF_0$  & \textbf{99\%} & \textbf{100\%} & \textbf{100\%} & \textbf{100\%} & \textbf{100\%} & \textbf{100\%} & 4\%\\
$APF_1$ & 98\% & 98\% & \textbf{100\%} & \textbf{100\%} & \textbf{100\%} & \textbf{100\%}  & 7\%\\
CF& 70\% & \textbf{100\%} & 86\% & 89\% & 88\% & 86\% &  9\%\\
ESF & 98\% & 4\% & \textbf{100\%} & \textbf{100\%} & \textbf{100\%} & 97\%  & 10\% \\ \hline
    \end{tabular}

\end{table}

Using different depth we could have obtained slightly different results. Since the purpose is to introduce the methodology for detecting the outliers we did not go that far in the analysis.

\section{Testing goodness of fit}
\label{sec:5}
%We want to test goodness of fit to a certain grem-grain model since it can give us an insight in whether there is any clustering or repulsiveness.
In this section, we present methods to test the goodness of fit of random set models considering different summary functions using the global envelope test.

Suppose we have a realisation of the random set and want to test whether it belongs to a particular random set model. We propose the following Monte Carlo method. First, we simulate $n$ realisations from the given random set model. For the chosen test summary function $T,$ we obtain its empirical estimate for the observed random set realisation, denoted by $T_1(r_i), i=1,\ldots,m$ and empirical estimates of the summary function for each simulated realisation, denoted by $T_j(r_i), j=2,\ldots,n+1, i=1,\ldots,m.$ In this way, we obtain $n+1$ discretised curves, which are the input data for the global envelope test.

For each $T_i$ $i=1,\ldots,n+1$ its extreme rank $R_i$ is calculated. The lower values of $R_i$ indicate that the corresponding curve $T_i$ is more extreme in a sample of curves.
We obtain the $p$-value of the test as the percentage of the curves whose ranks are more extreme than the rank of our observed test curve $T_1$:
\begin{equation*}
 p = \frac{1}{n+1}\left(\sum_{i=1}^{n+1} \mathbf{1}(R_i<R_1)\right).
\end{equation*}
The popularity of the global envelope test lies in the possibility of a graphical representation of the result. The construction of the graphical representation is as follows. For a given significance level $\alpha$, the rank $R_{\alpha}$ is selected as the the smallest rank for which
\begin{equation}
 \label{eq:pval}
 \sum\limits_{k=1}^{n+1}1( R_k<r_{(\alpha)})\geq \alpha(n+1).
\end{equation}

Then the envelope of all the curves whose rank is less than $R_{(\alpha)}$ is constructed.
The null hypothesis is not rejected if the observed vector lies completely within the envelope.
However, if the observed vector $T_1$ leaves the envelope at any point, the null hypothesis is rejected and the values $r$ for which $T_1(r)$ lies outside the envelope can provide the clue for the rejection.

In our simulation study, we used $50$ of realisations for each process. We used different test summary functions: $APF_0$, $APF_1$, extended empty space function (ESF), capacity functional (CF), support function of lift zonotop associated with the persistence diagram for features with dimension $0$ (denoted as $h_Z^0$) and support function of lift zonotop for features with dimension $1$ (denoted as $h_Z^1$).

The envelope test was performed in R with the function \verb|global_envelope_test| from the package \verb|GET| (see \cite{get_R}).

Table \ref{rezB} shows the percentage of rejected null hypothesis for each process and different test functions (the first number is the percentage of rejection for $p \leq 0.05$ and the second for $p \leq 0.1$), where the null hypothesis is a Boolean model. 

\begin{figure}[H]
    \centering
    \includegraphics[scale=0.7]{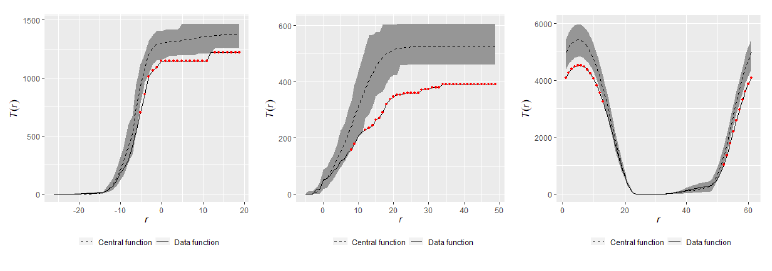}
    \caption{$95\%$ global envelope tests for different models and summary functions. First one is for recognising realisation of Matern cluster model among realisations of Boolean model using $APF_0$ as a test function. Second one if for recognising realisation of Matern cluster model among realisations of Boolean model using $APF_1$ as a test function. Third one is for  for recognising realisation of Cell model among realisations of DPP model using $h_Z^1$ as a test function (with fixed $\phi=\frac{\pi}{2}$).}
    \label{BvsMatClust}
\end{figure}

Figure \ref{BvsMatClust} shows examples of the graphical representation of the global envelope test. The first and second examples are used to detect the realisation of the Matern cluster model under the realisations of the Boolean model using $APF_0$ and $APF_1$ as test functions, respectively, and the third example is used to detect the realisation of theCell model among realisations of the DPP model using $h_Z^1$ (with fixed $\phi=\frac{\pi}{2}$) as test function. It is clear that the $APFs$ for the realisations of the Matern cluster model ”fall out” of the envelope at many points. When looking at $APF_1$, it becomes clear that the realisations of the Boolean model show a steeper increase in accumulated lifetimes than the realisations of the Matern cluster model, as loops with longer lifetimes are created earlier in the Boolean model. Smaller values of $APFs$ for larger meanages indicate clustering, which occurs in the Matern cluster model. The realisations of the Cell model have smaller values of $h_Z^1$ compared to the realisations of the DPP model because the realisations of the Cell model have slightly smaller empty space areas than the DPP model, as already mentioned in section \ref{liftzonoid}.

\setlength\tabcolsep{4pt}
\begin{table}[H]
\caption{Table showing percentage of rejected null hypothesis for each process and different test functions  (first number is percentage of rejection for $p\leq 0.05$ and second one is for $p \leq 0.1$). (Null hypothesis: Boolean)}
\label{rezB}
\begin{tabular}{lllllll} 
\hline
 & M & R & C & Be & Cell & DPP \\ \hline
 $h_Z^0$ & $4\%,\textcolor{lightgray}{42\%}$ & $\mathbf{100\%}$ & $\mathbf{100\%}$ & $\mathbf{100\%}$ & $40\%,\textcolor{lightgray}{74\%}$ & $60\%, \textcolor{lightgray}{88\%}$ \\ 
$h_Z^1$ & $34\%,\textcolor{lightgray}{52\%} $ & $42\%, \textcolor{lightgray}{54\%}$ & $68\%, \textcolor{lightgray}{92\%}$ & $48\%,\textcolor{lightgray}{60\%}$ & $\mathbf{100\%}$ & \multicolumn{1}{l}{$72\%, \textcolor{lightgray}{98\%}$}\\
$APF_0$ & $ \mathbf{88 \%},\textcolor{lightgray}{94\%}  $ & $\mathbf{100\%}$ & $\mathbf{100\%}$ & $68\%, \textcolor{lightgray}{100 \%}$ & $4\% ,\textcolor{lightgray}{64 \%} $ & $0\%,\textcolor{lightgray}{70 \%} $ \\ 
$APF_1$ & $66 \%, \textcolor{lightgray}{80\%}$ & $\mathbf{100\%}$ & $\mathbf{100\%}$ & $10\%,\textcolor{lightgray}{72\%} $ & $98\%, \textcolor{lightgray}{100 \%}$ & $50\%, \textcolor{lightgray}{96 \%}$ \\ 
ESF & $14\%, \textcolor{lightgray}{44\%}$ & $98\%,\textcolor{lightgray}{100\%} $ & $\mathbf{100\%}$ & $10\%, \textcolor{lightgray}{100\%}$ & $34\% ,\textcolor{lightgray}{100 \%} $ & $\mathbf{100 \%}$ \\ 
CF & $26\%, \textcolor{lightgray}{42\%}$ & $96\%$ , $\textcolor{lightgray}{100\%} $ & $\mathbf{100\%}$ & $46\%, \textcolor{lightgray}{88 \%}$ & $\mathbf{100} \%$ & $\mathbf{100\%}$
 \\ \hline
\end{tabular}
\end{table}

The results in the table \ref{rezB} are to be expected, considering that when exploring features of dimension $1$, we are actually trying to distinguish how voids and empty space occur in random set realisations. Since the Boolean model and the Boolean ellipse model have the same distribution of the seeds and the same configuration, i.e. the connected components are staggered in the same way, it is not surprising that $APF_1$ does not recognise many realisations of the Boolean ellipse model among the realisations of the Boolean model for $p \leq 0.05$. On the contrary, it works very well for the realisations of the other models because there are differences in the arrangement of the components due to repulsion or clustering. When using $APF_0$ and $APF_1$ as test functions for envelope tests, it outperforms the extended void function and the capacity functional in some cases, but not in all cases. Nevertheless, it is obvious that it can detect well whether a given realisation comes from the Boolean model or not.

Using the support function of the lift zonotop associated with the persistence diagram for features with dimension $0$ ($h_Z^0$) works very well in detecting realisations of all processes except the Matern cluster, and dimension $1$ ($h_Z^1$) works very well and is very powerful in detecting the Cell process and the DPP process.

The tables \ref{rez1}, \ref{rez2}, \ref{rez3}, \ref{rez4}, \ref{rez5} and \ref{rez6} show the percentage of rejected null hypothesis for each process and different test functions under different null hypotheses (the first number is the percentage of rejection for $p \leq 0.05$ and the second for $p \leq 0.1$).
\setlength\tabcolsep{4pt}
\begin{table}[H] 
\caption{Null hypothesis: Matern Cluster}
\label{rez1}
\begin{tabular}{lllllll} 
\hline
 & B & R & C & Be & Cell & DPP \\ \hline
 $h_Z^0$ & $18\%,\textcolor{lightgray}{42\%}$ & $\mathbf{100\%}$ & $\mathbf{100\%}$ & $\mathbf{100\%}$ & $88\%,\textcolor{lightgray}{100\%}$ & $\mathbf{100\%}$ \\ 
$h_Z^1$ & $0\%,\textcolor{lightgray}{86\%} $ & $0\%, \textcolor{lightgray}{100\%}$ & $64\%, \textcolor{lightgray}{100\%}$ & $0\%,\textcolor{lightgray}{100\%}$ & $0 \%,\textcolor{lightgray}{100\%}$ & \multicolumn{1}{l}{$0\%, \textcolor{lightgray}{100\%}$} \\
$APF_0$ & $ 66 \%,\textcolor{lightgray}{86\%}  $ & $\mathbf{100\%}$ & $74 \%, \textcolor{lightgray}{100\%} $ & $76\%, \textcolor{lightgray}{100 \%}$ & $80\% ,\textcolor{lightgray}{100 \%} $ & $\mathbf{100\%}$ \\ 
$APF_1$ & $\mathbf{68 \%}, \textcolor{lightgray}{72\%}$ & $\mathbf{100\%}$ & $80\%, \textcolor{lightgray}{100\%}$ & $14\%,\textcolor{lightgray}{22\%} $ & $98\%, \textcolor{lightgray}{100 \%}$ & $\mathbf{100\%}$ \\ 
ESF & $22\%, \textcolor{lightgray}{24\%}$ & $\mathbf{100\%} $ & $\mathbf{100\%}$ & $8\%, \textcolor{lightgray}{66\%}$ & $\mathbf{100 \%} $ & $\mathbf{100 \%}$ \\ 
CF & $10\%, \textcolor{lightgray}{20\%}$ & $\mathbf{100\%}$  & $\mathbf{100\%}$ & $6\%, \textcolor{lightgray}{100 \%}$ & $\mathbf{100} \%$ & $\mathbf{100\%}$ \\ 
 \hline
\end{tabular}
\end{table}

\setlength\tabcolsep{4pt}
\begin{table}[H]
\caption{Null hypothesis: Repulsive}
\label{rez2}
\begin{tabular}{lllllll} 
\hline
 & B & M & C & Be & Cell & DPP \\ \hline
 $h_Z^0$ &$\mathbf{100\%}  $ &$\mathbf{100\%}$ & $\mathbf{100\%} $ & $\mathbf{100\%}$ & $\mathbf{100\%} $ & $\mathbf{100\%}$  \\ 
$h_Z^1$ & $70\%,\textcolor{lightgray}{100\%} $ & $88\%, \textcolor{lightgray}{100\%}$ & $92\%, \textcolor{lightgray}{100\%}$ & $\mathbf{100\%}$ & $\mathbf{100\%} $& \multicolumn{1}{l}{$90\%,\textcolor{lightgray}{100\%}$} \\ 
$APF_0$ & $\mathbf{100\%}  $ &$\mathbf{100\%}$ & $\mathbf{100\%} $ & $\mathbf{100\%}$ & $\mathbf{100\%}$ & $\mathbf{100\%}$ \\ 
$APF_1$ & $\mathbf{100\%}  $ &$\mathbf{100\%}$ & $\mathbf{100\%}$ & $\mathbf{100\%}$ & $\mathbf{100\%} $ & $\mathbf{100\%}$  \\ 
ESF & $\mathbf{100\%}  $ &$\mathbf{100\%}$ & $\mathbf{100\%} $ & $\mathbf{100\%}$ & $90\%, \textcolor{lightgray}{100\%} $ & $16\%, \textcolor{lightgray}{100\%}$  \\ 
CF & $98\%, \textcolor{lightgray}{100\%}  $ &$\mathbf{100\%}$ & $\mathbf{100\%} $ & $\mathbf{100\%}$ & $18\%,\textcolor{lightgray}{86\%} $ & $8\%,\textcolor{lightgray}{82\%}$  \\ 
\hline
\end{tabular}
\end{table}

\setlength\tabcolsep{4pt}
\begin{table}[H]
\caption{Null hypothesis: Boolean Ellipse}
\label{rez3}
\begin{tabular}{lllllll} 
\hline
 & B & R & M & C & Cell & DPP \\ \hline
 $h_Z^0$ &$\mathbf{100\%}  $ &$\mathbf{100\%}$ & $\mathbf{98\%},\textcolor{lightgray}{100\%} $ & $68\%,\textcolor{lightgray}{100\%}$ & $\mathbf{100\%} $ & $\mathbf{100\%}$  \\ 
$h_Z^1$ & $26\%,\textcolor{lightgray}{66\%} $ & $\mathbf{100\%}$ & $68\%, \textcolor{lightgray}{94\%}$ & $\mathbf{100\%}$ & $98\%,\textcolor{lightgray}{100\%} $& \multicolumn{1}{l}{$94\%,\textcolor{lightgray}{100\%}$} \\
$APF_0$ & $26\%,\textcolor{lightgray}{100\%}  $ &$\mathbf{100\%}$ & $96\%, \textcolor{lightgray}{100\%} $ & $\mathbf{100\%}$ & $40\%,\textcolor{lightgray}{100\%} $ & $82\%,\textcolor{lightgray}{100\%}$ \\ 
$APF_1$ & $30\%,\textcolor{lightgray}{64\%} $ &$\mathbf{100\%}$ & $48\%,\textcolor{lightgray}{86\%} $ & $98\%, \textcolor{lightgray}{100\%}$ & $98\%, \textcolor{lightgray}{100\%} $ & $\mathbf{100\%}$  \\ 
ESF & $10\%,\textcolor{lightgray}{100\%}  $ &$\mathbf{100\%}$ & $16\%,\textcolor{lightgray}{96\%} $ & $\mathbf{100\%}$ & $98\%, \textcolor{lightgray}{100\%} $ & $\mathbf{100\%}$  \\ 
CF & $50\%, \textcolor{lightgray}{96\%}  $ &$\mathbf{100\%}$ & $18\%,\textcolor{lightgray}{88\%} $ & $\mathbf{100\%}$ & $\mathbf{100\%} $ & $\mathbf{100\%}$  \\ 
 \hline
\end{tabular}
\end{table}

\setlength\tabcolsep{4pt}
\begin{table}[H]
\caption{Null hypothesis: Cluster}
\label{rez4}
\begin{tabular}{lllllll} 
\hline
 & B & R & M & Be & Cell & DPP \\ \hline
 $h_Z^0$ &$\mathbf{100\%}  $ &$\mathbf{100\%}$ & $\mathbf{100\%} $ & $44\%,\textcolor{lightgray}{76\%}$ & $\mathbf{100\%} $ & $\mathbf{100\%}$  \\ 
$h_Z^1$ & $\mathbf{100\%} $ & $\mathbf{100\%}$ & $90\%, \textcolor{lightgray}{98\%}$ & $98\%,\textcolor{lightgray}{100\%}$ & $\mathbf{100\%} $& \multicolumn{1}{l}{$\mathbf{100\%}$} \\
$APF_0$ & $\mathbf{100\%}  $ &$\mathbf{100\%}$ & $70, \textcolor{lightgray}{100\%} $ & $\mathbf{100\%}$ & $\mathbf{100\%} $ & $\mathbf{100\%}$ \\ 
$APF_1$ & $\mathbf{100\%}  $ &$\mathbf{100\%}$ & $44\%,\textcolor{lightgray}{100\%} $ & $98\%,\textcolor{lightgray}{100\%}$ & $\mathbf{100\%} $ & $\mathbf{100\%}$  \\ 
ESF & $\mathbf{100\%}  $ &$\mathbf{100\%}$ & $\mathbf{100\%} $ & $\mathbf{100\%}$ & $\mathbf{100\%} $ & $\mathbf{100\%}$  \\ 
CF & $\mathbf{100\%}$ &$\mathbf{100\%}$ & $\mathbf{100\%}$ & $\mathbf{100\%}$ & $\mathbf{100\%}$ & $\mathbf{100\%}$  \\ 
 \hline
\end{tabular}
\end{table}

\setlength\tabcolsep{4pt}
\begin{table}[H]
\caption{Null hypothesis: Cell process}
\label{rez5}
\begin{tabular}{lllllll} 
\hline
 & B & R & M & Be & C & DPP \\ \hline
 $h_Z^0$ &$94\%,\textcolor{lightgray}{96\%}  $ &$\mathbf{100\%}$ & $\mathbf{100\%}$ & $\mathbf{100\%}$ & $\mathbf{100\%} $ & $12\%,\textcolor{lightgray}{38\%}$  \\ 
$h_Z^1$ & $\mathbf{100\%} $ & $\mathbf{100\%}$ & $\mathbf{100\%}$ & $\mathbf{100\%}$ & $\mathbf{100\%} $& \multicolumn{1}{l}{$\mathbf{60\%},\textcolor{lightgray}{94\%}$} \\
$APF_0$ & $10\%,\textcolor{lightgray}{74\%}  $ &$\mathbf{100\%}$ & $94\%, \textcolor{lightgray}{100\%} $ & $14\%, \textcolor{lightgray}{100\%}$ & $\mathbf{100\%}$ & $18\%,\textcolor{lightgray}{40\%}$ \\ 
$APF_1$ & $34\%,\textcolor{lightgray}{100\%} $ &$\mathbf{100\%}$ & $96\%,\textcolor{lightgray}{100\%} $ & $88\%, \textcolor{lightgray}{100\%}$ & $\mathbf{100\%}$ & $18\%,\textcolor{lightgray}{44\%}$  \\ 
ESF & $96\%,\textcolor{lightgray}{100\%}  $ & $90\%,\textcolor{lightgray}{100\%}$ & $\mathbf{100\%} $ & $\mathbf{100\%}$ & $\mathbf{100\%}$ & $46\%,\textcolor{lightgray}{78\%}$  \\ 
CF & $\mathbf{100\%}  $ &$22\%,\textcolor{lightgray}{88\%}$ & $\mathbf{100\%} $ & $\mathbf{100\%} $ & $\mathbf{100\%} $ & $0\%, \textcolor{lightgray}{0\%} $  \\ 
 \hline
\end{tabular}
\end{table}

\setlength\tabcolsep{4pt}
\begin{table}[H]
\caption{Null hypothesis: DPP}
\label{rez6}
\begin{tabular}{lllllll} 
\hline
 & B & R & M & Be & C & Cell \\ \hline
 $h_Z^0$ &$98\%,\textcolor{lightgray}{100\%}  $ &$\mathbf{100\%}$ & $\mathbf{100\%}$ & $\mathbf{100\%}$ & $\mathbf{100\%} $ & $42\%,\textcolor{lightgray}{56\%}$  \\ 
$h_Z^1$ & $\mathbf{100\%} $ & $\mathbf{100\%}$ & $\mathbf{100\%}$ & $\mathbf{100\%}$ & $\mathbf{100\%} $& \multicolumn{1}{l}{$10\%,\textcolor{lightgray}{56\%}$} \\
$APF_0$ & $64\%,\textcolor{lightgray}{100\%}  $ &$\mathbf{100\%}$ & $\mathbf{100\%} $ & $82\%, \textcolor{lightgray}{100\%}$ & $\mathbf{100\%}$ & $36\%,\textcolor{lightgray}{54\%}$ \\ 
$APF_1$ & $92\%,\textcolor{lightgray}{98\%} $ &$\mathbf{100\%}$ & $\mathbf{100\%} $ & $\mathbf{100\%}$ & $\mathbf{100\%}$ & $26\%,\textcolor{lightgray}{28\%}$  \\ 
ESF & $\mathbf{100\%}  $ & $84\%,\textcolor{lightgray}{96\%}$ & $\mathbf{100\%} $ & $\mathbf{100\%}$ & $\mathbf{100\%} $ & $\mathbf{72\%},\textcolor{lightgray}{72\%}$  \\ 
CF & $\mathbf{100\%} $ &$34\%,\textcolor{lightgray}{100\%}$ & $\mathbf{100\%}$ & $\mathbf{100\%} $ & $\mathbf{100\%}  $ & $8\%, \textcolor{lightgray}{20\%} $  \\ 
 \hline
\end{tabular}
\end{table}

\section{Discussion}
The main objective of the paper was to leverage the tools of the topological data analysis for statistical goodness-of-fit testing and outlier detection in samples of random sets with a particular focus on germ-grain random set models. Considering APFs and the newly introduced support function of the lift zonotop of PD as summary functions, we investigated the possibility of detecting various interactions among the grains, such as different types of clustering or repulsion tendencies.
We have proposed methods for detecting outliers and testing the goodness-of-fit of random set models based on the summary functions.

The simulation study has shown that the APFs and the support function of the lift zonotop of PD can detect the outliers and the differences between the germ-grain models better than well-known summary functions such as the capacity functional or the extended empty space function.

Since our goodness of fit test procedure uses Monte Carlo simulations, which are computationally expensive, our future work will focus on proving the CLT for APFs and support functions of the lift zonotop to speed up the testing procedure.

We also plan to apply the method to some real data and investigate the possibility of using TDA methods for the purpose of the classification of random sets.

\section*{Funding}
The first author was supported in part by  the Croatian-Swiss Research Program of the Croatian Science Foundation and the Swiss National Science Foundation: project number IZHRZ0\_180549.

\section*{Supplement: Detailed percentages of the types of outliers (1st, 2nd and 3rd order) for all combination of models and added realisations when the simulated sample of 30 realisations comes from Boolean, repulsive, cluster, Boolean-ellipse, Cell and DDP model.}
\label{sup}

\begin{table}[H]
 \caption{Boolean :Outlier types}
\label{tab:out_type_B}
\centering
\begin{tabular}{|c|ccc|ccc|ccc|}
\hline
 &  \multicolumn{3}{c}{R}  \vline & \multicolumn{3}{c}{C} \vline  & \multicolumn{3}{c}{Be} \vline  \\
 \hline
& 1  & 2 & 3&  1   & 2 & 3 &  1   & 2 & 3   \\ \hline
  $h_Z^0$  & 79\%  & 9\% & 9\%  &100\% &  0\% &  0\% &   80\% & 1\% & 16\%\\
 $h_Z^1$ & 48\% &  0\% & 12\% & 76\% & 3\% & 4\%  & 63\% &  1\% & 1\% \\
  $APF_0$ & 99\%  & 0\% & 1\%  & 100\%  & 0\% &  0\% &    7\% & 21\% &  6\%\\
 $APF_1$ & 96\% & 0\%  & 4\% & 100\% &   0\% &   0\% &  11\% & 15\% & 4\%  \\
 CF & 58\% & 9\% & 9\% & 100\%  & 0\% &  0\% &  53\% & 20\% &  6\%\\
 ESF & 4\% & 1\% & 12\% &  100\%  & 0\%  & 0\% &     24\% &26\% &  2\% \\
 \hline
  &  \multicolumn{3}{c}{M} \vline& \multicolumn{3}{c}{Cell} \vline  & \multicolumn{3}{c}{DPP} \vline\\ \hline
   & 1   & 2 & 3 &  1   & 2 & 3 &  1   & 2 & 3  \\\hline
   $h_Z^0$  & 1\% & 0\% & 4\% & 73\% &  2\% & 16\% & 85\% & 2\% &  5\% \\
  $h_Z^1$ & 25\% & 1\% & 5\% &   71\% & 1\% & 5\% &  55\% & 1\% & 3\% \\
  $APF_0$ & 57\% &  2\%  & 8\% & 38\% & 0\% & 9\% & 90\% & 2\% & 1\%\\
  $APF_1$  &  32\% &  2\% &  8\% &72\% & 0\% & 9\% & 91\% &  0\% & 2\%\\
  CF & 53\% & 20\% & 5\% & 53\% & 20\% &  6\% & 53\% & 20\% & 2\% \\
  ESF &18\% & 14\% & 5\% & 0\% & 3\% & 5\% & 5\% & 6\%  & 28\% \\
  \hline
\end{tabular}
\end{table}

\begin{table}[H]
 \caption{Repulsive :Outlier types}
\label{tab:out_type_R}
\centering
\begin{tabular}{|c|ccc|ccc|ccc|}
\hline
 &  \multicolumn{3}{c}{B} \vline & \multicolumn{3}{c}{C} \vline  & \multicolumn{3}{c}{Be} \vline \\
 \hline
& 1   & 2 & 3 &  1   & 2 & 3 & 1   & 2 & 3 \\
\hline
 $h_Z^0$  &99\%  & 0\% & 0\% & 100\% & 0\% &  0\% &  100\% & 0\% & 0\% \\
 
 $h_Z^1$ & 70\% & 1\% & 11\% & 91\% &  2\% & 4\% & 55\% & 4\% & 5\% \\
  $APF_0$ & 100\%  & 0\% & 0\%  &  100\%  & 0\% &  0\% &     100\% & 0\% &  0\%  \\
 $APF_1$ & 100\% & 0\%  & 0\% &   100\% &   0\% &   0\% &  100\% & 0\% & 0\% \\

 CF & 51\% & 6\% & 14\% & 100\%  & 0\% &  0\% &   52\%  & 23\% & 2\% \\

 ESF & 94\% & 3\% &  2\%  & 100\%  & 0\%  & 0\%  &  100\% & 0\% & 0\% \\
 \hline
 &  \multicolumn{3}{c}{M} \vline & \multicolumn{3}{c}{Cell} \vline & \multicolumn{3}{c}{DPP} \vline  \\
 \hline
 & 1   & 2 & 3& 1   & 2 & 3 & 1   & 2 & 3 \\ \hline
 $h_Z^0$ &  99\% & 0\% & 1\% &  100\% & 0\% & 0\% & 94\% & 0\% & 5\% \\
  $h_Z^1$ &  86\% & 1\% & 5\% &  100\% & 0\% & 0\% &  100\% &  0\% &  0\% \\
 $APF_0$ &  100\%  & 0\% &  0\% &   100\% & 0\% & 0\% &  100\%  & 0\% &  0\%  \\
 $APF_1$ &  100\%  & 0\% &  0\% &100\% & 0\% & 0\% &  100\% &  0\%  & 0\% \\
 CF   & 52\% & 23\% & 6\% & 52\% & 23\% & 4\% & 52\% & 23\% & 4\% \\
 ESF  &97\% & 2\%  & 1\% &   85\% &  8\% & 2\% & 13\% & 1\%& 4\% \\
 \hline
\end{tabular}
\end{table}

\begin{table}[H]
 \caption{Cluster :Outlier types}
\label{tab:out_type_C}
\centering
\begin{tabular}{|c|ccc|ccc|ccc|}
\hline
 &  \multicolumn{3}{c}{R} \vline & \multicolumn{3}{c}{B}  \vline & \multicolumn{3}{c}{Be}  \vline \\ \hline
& 1   & 2 & 3 &  1   & 2 & 3 &  1   & 2 & 3 \\
\hline
  SLZ0  &64\% & 36\% & 0\%  &98\% &  2\% &  0\% & 69\% & 31\% & 0\%\\
 
 SLZ1 & 23\% & 14\% & 27\% & 3\% & 1\% & 8\% &  26\% & 3\% & 14\%\\
  $APF_0$ & 99\% & 1\% & 0\% & 98\%  & 2\% &  0\% &     100\% & 0\% &  0\% \\
 $APF_1$ & 100\% & 0\%  & 0\% &   99\% &   0\% &  1\%  & 94\% & 0\% & 6\% \\

 CF & 51\% & 6\% & 16\% & 100\%  & 0\% &  0\%  & 64\% & 13\% & 1\% \\

 ESF & 100\% & 0\%  & 0\%  & 100\%  & 0\%  & 0\% &    100\% & 0\% & 0\% \\
 \hline
 &  \multicolumn{3}{c}{M}  \vline & \multicolumn{3}{c}{Cell}  \vline  & \multicolumn{3}{c}{DPP}  \vline \\
 \hline
 & 1   & 2 & 3 &  1   & 2 & 3 &  1   & 2 & 3 \\ \hline
 $h_Z^0$  & 99\% & 0\% & 1\% & 100\% & 0\% & 0\%  & 100\% & 0\% & 0\%  \\
 $h_Z^1$ & 6\% & 2\% & 6\% & 57\% & 7\% & 12\% &   22\% & 4\% & 14\%  \\
 $APF_0$  &  99\%  & 1\% &  0\% & 100\% & 0\% & 0\% &    99\% & 1\% & 0\% \\
 $APF_1$ &  95\%  & 0\% &  3\% &100\% & 0\% & 0\%  & 99\% &  0\%  & 1\% \\
 CF & 64\% & 13\% &  3\% &64\% &13\% &  4\% &  64\% & 13\% &  4\%  \\
 ESF & 100\% & 0\% &  0\% &100\% &0\% &  0\% &  100\% & 0\% &  0\%  \\
 \hline
\end{tabular}
\end{table}

\begin{table}[H]
 \caption{Boolean ellipse :Outlier types}
\label{tab:out_type_Be}
\centering
\begin{tabular}{|c|ccc|ccc|ccc|}
\hline
 &  \multicolumn{3}{c}{R} \vline & \multicolumn{3}{c}{B} \vline & \multicolumn{3}{c}{C} \vline \\ \hline
& 1   & 2 & 3 & 1   & 2 & 3 & 1   & 2 & 3 \\ \hline
  $h_Z^0$ & 87\%  &2\%  &6\%  &100\%  &0\%  &0\%    &98\%  &2\%  &0\% \\
$h_Z^1$ &78\%  &2\%  &3\%    &27\%  &2\%  &7\%    &89\%  &3\%  &4\% \\
$APF_0$ &35\%  &36\%  &0\%   & 100\%  &0\%  &0\%  & 100\%  &0\%  &0\% \\
$APF_1$ & 20\%  &28\%  &1\%   &100\%  &0\%  &0\%   & 100\%  &0\%  &0\% \\
CF&  50\%  &9\%  &16\% & 100\%  &0\%  &0\%   &63\%  &16\%  &4\% \\
ESF & 17\%  &6\%  &14\%   &66\%  &5\%  &11\%  & 100\%  &0\%  &0\% \\
\hline
&  \multicolumn{3}{c}{M} \vline & \multicolumn{3}{c}{Cell} \vline  & \multicolumn{3}{c}{DPP} \vline \\
 \hline
 &  1   & 2 & 3 & 1   & 2 & 3 & 1   & 2 & 3 \\ \hline
 $h_Z^0$ &78\% &4\% &18\%  & 100\% &0\% &0\% &99\% & 0\% & 1\%\\
 $h_Z^1$  &85\% &0\% &3\%  & 100\% &0\% &0\% &99\% & 1\% & 0\% \\
 $APF_0$  &69\% &23\% &0\%  &83\% &9\% &0\% & 100\%& 0\% & 0\% \\
 $APF_1$ &44\% &23\% & 4\% &99\% &0\% &1\%  &100\% & 0\% & 0\%\\
 CF & 63\% &16\% & 1\% &63\% &16\% &2\%  &63\% &16\% &  3\% \\
 ESF& 14\% &10\% &7\% &47\% &4\% &17\% &61\%& 8\% &17\% \\
\hline
 \end{tabular}
\end{table}

\begin{table}[H]
 \caption{Matern :Outlier types}
\label{tab:out_type_M}
\centering
\begin{tabular}{|c|ccc|ccc|ccc|}
\hline
 &  \multicolumn{3}{c}{R} \vline & \multicolumn{3}{c}{B}  \vline  & \multicolumn{3}{c}{C}  \vline  \\ \hline
& 1   & 2 & 3& 1   & 2 & 3 & 1   & 2 & 3 \\ \hline
  $h_Z^0$ & 2\% & 2\% & 2\% &    98\% & 2\% & 0\% & 100\% & 0\% & 0\% \\
$h_Z^1$ & 2\% & 0\% & 0\% &   69\% & 1\% & 5\% &  42\% & 0\% & 11\% \\
$APF_0$ & 48\% & 6\% & 4\% &   99\% & 0\% & 1\% &   98\% & 1\% &  1\% \\
$APF_1$ & 25\% & 0\%  &  14\% & 100\% & 0\% & 0\% &  94\% & 0\% & 5\%\\
CF & 48\% & 5\%  &  20\%  & 100\% & 0\% & 0\% &    50\% & 33\% & 2\% \\
ESF & 2\% & 1\% & 2\% &   27\% & 1\%  &  17\%  & 99\% & 1\% & 0\% \\
 \hline
 &  \multicolumn{3}{c}{Be}  \vline & \multicolumn{3}{c}{Cell}  \vline & \multicolumn{3}{c}{DPP}  \vline \\ \hline
 &1   & 2 & 3& 1   & 2 & 3 & 1   & 2 & 3 \\ \hline
$h_Z^0$  &69\% & 0\% & 18\% & 100\% & 0\% & 0\% & 75\% & 3\% & 9\% \\
$h_Z^1$ &79\% & 3\% & 1\%  & 35\% & 3\% & 12\%  & 51\% & 5\% & 7\% \\
$APF_0$ & 44\% & 13\% & 1\% & 99\% & 1\% & 0\% & 85\% & 6\% & 2\%\\
$APF_1$  & 9\% & 6\% & 4\%  & 98\% & 0\% & 1\% & 99\% & 0\% & 1\% \\ 
CF & 50\% & 33\% & 1\% &  50\% & 33\% & 3\%  & 50\% & 33\% & 0\% \\
ESF &3\% & 12\% & 3\% & 100\% & 0\% & 0\%  & 22\% & 1\% & 8\% \\
 \hline
\end{tabular}
\end{table}

\begin{table}[H]
 \caption{Cell :Outlier types}
\label{tab:out_type_Cell}
\centering
\begin{tabular}{|c|ccc|ccc|ccc|}
\hline
 &  \multicolumn{3}{c}{R} \vline & \multicolumn{3}{c}{B} \vline & \multicolumn{3}{c}{C} \vline \\ \hline
& 1   & 2 & 3 &  1   & 2 & 3 & 1   & 2 & 3\\ \hline
  $h_Z^0$ & 89\% & 0\% & 6\% & 100\% & 0\% & 0\%  & 100\%  &  0\%  &  0\%  \\
$h_Z^1$ & 68\% & 8\% & 7\% & 100\% & 0\% & 0\% & 88\%  &  4\%  &  2\%   \\
$APF_0$ & 37\% & 15\% & 2\% & 99\% & 0\% & 1\% & 100\%  &  0\%  &  0\%\\
$APF_1$ & 87\% & 6\% & 0\%  & 98\% & 0\% & 2\%  & 100\%  &  0\%  &  0\% \\
CF & 44\% & 14\% & 16\% & 100\% & 0\% & 0\% &  57\% & 21\%  &  4\%\\
ESF & 48\% & 14\% & 10\% & 0\%  & 1\% & 25\% &  100\%  &  0\%  &  0\% \\
 \hline
  &  \multicolumn{3}{c}{Be} \vline & \multicolumn{3}{c}{M} \vline & \multicolumn{3}{c}{DPP} \vline  \\
 \hline
 & 1   & 2 & 3 &  1   & 2 & 3 & 1   & 2 & 3  \\ \hline
 $h_Z^0$ & 100\% & 0\% & 0\% & 100\% & 0\% & 0\%  &  94\%  &  0\%  &  0\%  \\
 $h_Z^1$ &98\% & 2\% & 0\% &   83\% & 8\% & 2\%  &  73\%  &  6\%  &  5\%  \\
 $APF_0$  & 53\% &  26\% & 1\% & 100\% & 0\% & 0\% &  95\%  &  1\%  &  1\%  \\
 $APF_1$ & 97\% & 0\% & 0\% &  100\% & 0\% & 0\% &   99\%  &  0\%  &  1\%  \\
 CF   &57\% &  21\% & 1\% &   57\% &  21\% & 2\% &  57\% & 21\%  &  5\%  \\
 ESF &72\% &  13\% & 7\% &  100\% & 0\% & 0\%  &  72\% & 10\%  &  0\%  \\
 \hline
\end{tabular}
\end{table}

\begin{table}[H]
 \caption{DPP:Outlier types}
\label{tab:out_type_DPP}
\centering
\begin{tabular}{|c|ccc|ccc|ccc|}
\hline
 &  \multicolumn{3}{c}{R} \vline & \multicolumn{3}{c}{B}  \vline  & \multicolumn{3}{c}{C}  \vline  \\ \hline
& 1   & 2 & 3 & 1   & 2 & 3 & 1   & 2 & 3\\ \hline
$h_Z^0$ & 95\% & 0\% & 2\% & 99\% & 0\% & 1\%  & 100\%  &  0\%  &  0\%\\
$h_Z^1$ & 73\% & 6\% & 4\% & 100\% & 0\% & 0\% &  88\%  &  7\%  &  4\%\\  
$APF_0$ & 98\% & 1\% & 0\% & 100\% & 0\% & 0\%  & 100\%  &  0\%  &  0\%  \\
$APF_1$ & 97\% & 1\% & 0\% &  96\% & 0\% & 2\% & 100\%  &  0\%  &  0\%  \\
CF & 44\% & 8\% & 18\% &  100\% & 0\% & 0\% & 61\% & 25\%  &  0\% \\
ESF &76\% & 21\% & 1\% & 0\%  & 3\% & 1\% &  100\%  &  0\%  &  0\%  \\
\hline
 &  \multicolumn{3}{c}{Be}  \vline  & \multicolumn{3}{c}{M}  \vline  & \multicolumn{3}{c}{Cell}  \vline  \\
 \hline
 & 1   & 2 & 3 & 1   & 2 & 3 & 1   & 2 & 3  \\ \hline
$h_Z^0$    & 100\% & 0\% & 0\%  & 100\% & 0\% & 0\% &  94\%  &  0\%  &  3\%  \\
$h_Z^1$& 100\% & 0\% & 0\% &  88\% & 5\% & 2\% &  61\%  &  3\%  &  4\%  \\
$APF_0$ & 99\% & 1\% & 0\%  & 100\% & 0\% & 0\%  & 100\%  &  0\%  &  0\%  \\
$APF_1$ & 100\% & 0\% & 0\%  & 100\% & 0\% & 0\%  & 100\%  &  0\%  &  0\%  \\
CF & 61\% &  25\% & 3\% &  61\% &  25\% & 2\% &  61\% & 25\%  &  0\%  \\
ESF & 91\% & 9\% & 0\% & 100\% & 0\% & 0\% &   86\%  &  8\%  &  3\%  \\

 \hline

 \end{tabular}
\end{table}

\end{document}